\documentclass[aps,twocolumn,floats,prd,nofootinbib,10pt,longbibliography,superscriptaddress]{revtex4-1}

\usepackage{comment}
\usepackage[dvips]{graphicx} %
\usepackage{graphicx,amsmath,amsfonts,amssymb,slashed,float,hyperref}
\usepackage[normalem]{ulem}
\usepackage{bbold,wasysym}
\usepackage{graphicx}
\usepackage{array,multirow}
\usepackage[utf8]{inputenc}

\usepackage[usenames,dvipsnames]{xcolor} 

\usepackage{soul}

\definecolor{RoyalBlue}{rgb}{0.25,.41,.88}
\definecolor{celestialblue}{rgb}{0.29, 0.59, 0.82}

\setstcolor{Blue}
\def\mnras{MNRAS}
\def\apjs{ApJS}

\def\AB#1{\textcolor{Magenta}{AB}}

\newcommand{\LCDM}{$\Lambda$CDM}

\begin{document}

\title{Dark Energy at early times and ACT: a larger Hubble constant without late-time priors}

\author{Vivian Poulin}
\affiliation{Laboratoire Univers \& Particules de Montpellier (LUPM),
CNRS \& Universit\'{e} de Montpellier (UMR-5299),
Place Eug\`{e}ne Bataillon, F-34095 Montpellier Cedex 05, France}
\author{Tristan L.~Smith}
\affiliation{Department of Physics and Astronomy, Swarthmore College, Swarthmore, PA 19081, USA}
\author{Alexa Bartlett}
\affiliation{Department of Physics and Astronomy, Swarthmore College, Swarthmore, PA 19081, USA}

\begin{abstract}
Recent observations using the Atacama Cosmology Telescope (ACT) have provided ground-based cosmic microwave background (CMB) maps with higher angular resolution than the {\it Planck} satellite. These have the potential to put interesting constraints on models resolving the `Hubble tension'. In this paper we fit two models of Early Dark Energy (EDE) (an increase in the expansion rate around matter/radiation equality) to the combination of ACT data with large-scale measurements of the CMB either from the WMAP or the {\it Planck} satellite (including lensing), along with  measurements of the baryon acoustic oscillations and uncalibrated supernovae luminosity distance.  We study a phenomenological axion-like potential (`axEDE') and a scalar field experiencing a first-order phase-transition (`NEDE').  We find that for both models the `{\it Planck}-free' analysis yields non-zero EDE at $\gtrsim 2\sigma$ and an increased value for $H_0
\sim 70-74$ km/s/Mpc, compatible with local measurements,  without the inclusion of any prior on $H_0$. On the other hand, the inclusion of {\it Planck} data restricts the EDE contribution to an upper-limit only at 95\% C.L. For axEDE, the combination of {\it Planck} and ACT leads to constraints 30\% weaker than with {\it Planck} alone, and there is no residual Hubble tension. On the other hand, NEDE is more strongly constrained in a {\it Planck}+ACT analysis, and the Hubble tension remains at $\sim 3\sigma$, illustrating the ability for CMB data to distinguish between different EDE models. We further explore the apparent inconsistency between the {\it Planck} and ACT data and find that it comes (mostly) from a slight tension between the temperature power spectrum at multipoles around $\sim 1000$ and $\sim 1500$.  Finally, through a mock analysis of ACT data, we demonstrate that the preference for EDE is not driven by a lack of information at high-$\ell$ when removing {\it Planck} data, and that a \LCDM\ fit to the fiducial EDE cosmology results in a significant bias on $\{H_0,\omega_{\rm cdm}\}$. More accurate measurements of the TT CMB power spectra above $\ell\sim 2500$ and EE between $\ell \sim 300-500$ will play a crucial role in differentiating between EDE models.

\end{abstract}
\date{\today}

\maketitle

\section{Introduction}
\label{sec:into}

Over the past several years, the standard cosmological model, \LCDM{}, has come under increased scrutiny as measurements of the late-time expansion history of the Universe~\cite{Scolnic:2017caz}, the cosmic microwave background (CMB)~\cite{Planck:2019nip}, and large-scale structure (LSS)---such as the clustering of galaxies~\cite{Alam:2016hwk,Abbott:2017wau,Hildebrandt:2018yau,eBOSS_cosmo}---have improved. 
Observations have spurred recent tensions within \LCDM{}, related to the Hubble constant $H_0 = 100 h$ km/s/Mpc~\cite{Verde:2019ivm} and the parameter combination $S_8 \equiv \sigma_8(\Omega_{\rm m}/0.3)^{0.5}$~\cite{Joudaki:2019pmv} (where $\Omega_{\rm m}$ is the total matter relic density and $\sigma_8$ is the variance of matter perturbations within 8 Mpc/$h$ today), reaching the $\sim 4-5\sigma$ \cite{Riess:2020fzl,Soltis:2020gpl} and $2-3\sigma$ level \cite{Joudaki:2019pmv,Heymans:2020gsg,DES:2021wwk}, respectively. While both of these tensions may be the result of systematic uncertainties, and not all measurements lead to the same tension\footnote{See Refs.~\cite{Freedman:2021ahq,Anand:2021sum} for recent interesting discussions about potential systematic offset between the two most common methods to calibrate SN1a, namely Cepheid variable stars and the `tip of the red giant branch'.} \cite{Freedman:2019jwv,Freedman:2020dne}, numerous models have been suggested as potential resolution to these tensions (see e.g.~Ref.~\cite{DiValentino:2021izs,Schoneberg:2021qvd} for recent reviews). Yet, no model is able to resolve both tensions simultaneously \cite{Jedamzik:2020zmd,Schoneberg:2021qvd}. In this work we focus on models which may resolve the Hubble tension.

We specifically focus on models of `Early Dark Energy' (EDE), one of the most promising ways to resolve the Hubble tension \cite{Schoneberg:2021qvd}. These models posit an additional energy density which increases the expansion rate before recombination and then dilutes faster than radiation. EDE is usually modeled through a scalar-field with negligible energy density until a critical redshift (around matter-radiation equality) after which the field again becomes negligible. When fit to {\it Planck}, baryon acoustic oscillations (BAO) and supernova type Ia (SN1a) Pantheon data, these models have been shown to reduce the Hubble tension to the $1.5\sigma$ level \cite{Schoneberg:2021qvd}. 

While {\it Planck} alone does not favor EDE, such models leave an impact in the CMB power spectra at $\ell\gtrsim 500$ that are not fully degenerate with $\Lambda$CDM, providing a way to detect (or exclude) EDE with high accuracy CMB measurements \cite{Smith:2019ihp}. Recently, the ACT collaboration \cite{Aiola:2020azj} has produced ground-based CMB maps with higher angular resolution than the {\it Planck} satellite \cite{Planck:2019nip} that have the potential to put interesting constraints on EDE models. The goal of this article is to fit two models of EDE from the recent literature to the combination of ACT data with large-scale measurements of the CMB either from the WMAP or {\it Planck}, along with  measurements of the BAO and uncalibrated SN1a from Pantheon data: the phenomenological axion-like `Early Dark Energy' (axEDE) model from Ref.~\cite{Smith:2019ihp} and the `new' EDE (NEDE) model from Ref.~\cite{Niedermann:2019olb}.

We show that ACT data (with and without WMAP) prefers a non-zero EDE contribution at $\gtrsim 2\sigma$, regardless of the model, {\em without the inclusion of any prior on $H_0$} leading to no residual tension between WMAP+ACT and SH0ES. This is in contrast with the results from {\it Planck}, which have been shown to place an upper limit\footnote{From hereon we quote two-sided constraints at 68\% C.L. and one-sided lower and upper limit at 95\% C.L..} on the fraction at the critical redshift $f_{\rm axEDE}(z_c) \lesssim 0.089$ and $f_{\rm NEDE}(z_*) \lesssim 0.116$.
In a conservative approach, we combine a restricted version of ACT \cite{Aiola:2020azj} with {\it Planck}, finding a {\em weaker} upper limit than from {\it Planck}-only, $f_{\rm axEDE}(z_c)<0.110$. We find that there is no residual tension with SH0ES ($0.3\sigma$ agreement), a remarkable improvement over $\Lambda$CDM for which the tension is $4.5\sigma$. However, the NEDE model is more strongly constrained with these data, yielding $f_{\rm NEDE}(z_*)<0.082$ leaving a residual $\sim 3\sigma$ tension. This is an example of how the combination of Planck and ACT can break degeneracies between different axEDE models.

We then investigate the origin of the difference between the {\it Planck} and WMAP+ACT results. We perform an MCMC analysis of {\it Planck} and ACT, now restricting the {\it Planck} $\ell$-range to $\ell < 1060$ (which roughly matches the range of $\ell$ covered by WMAP), and find similar results as the WMAP+ACT analysis. We identify that the EDE cosmologies favored by ACT data lead to a TT power spectrum that is systematically higher than that measured by {\it Planck} at $\ell > 1000$, with most notable differences at $\ell\sim 1000$ and $\ell \sim 1500$. Moreover, while the axEDE preference is relatively unaffected by {\it Planck} polarization data, the NEDE model is constrained by this dataset. 

From the analysis of mock ACT data (focusing on axEDE) we find that the preference for a non-zero EDE contribution does not come from a lack of information when leaving out the {\it Planck} data, which might bias the MCMC posterior distributions. Assuming the $\Lambda$CDM best-fit cosmology as the fiducial model, the reconstructed axEDE posteriors are compatible with $f_{\rm axEDE}(z_c)=0$ at $1\sigma$, in stark contrast with the results of the real data analysis.  On the other hand, with an axEDE best-fit cosmology as the fiducial, we find that the reconstructed posteriors for $\Lambda$CDM \emph{agree} with the real analysis and are up to $7\sigma$ biased with respect to their fiducial (axEDE) values. 

\begin{figure*}
    \centering
\includegraphics[scale=0.7]{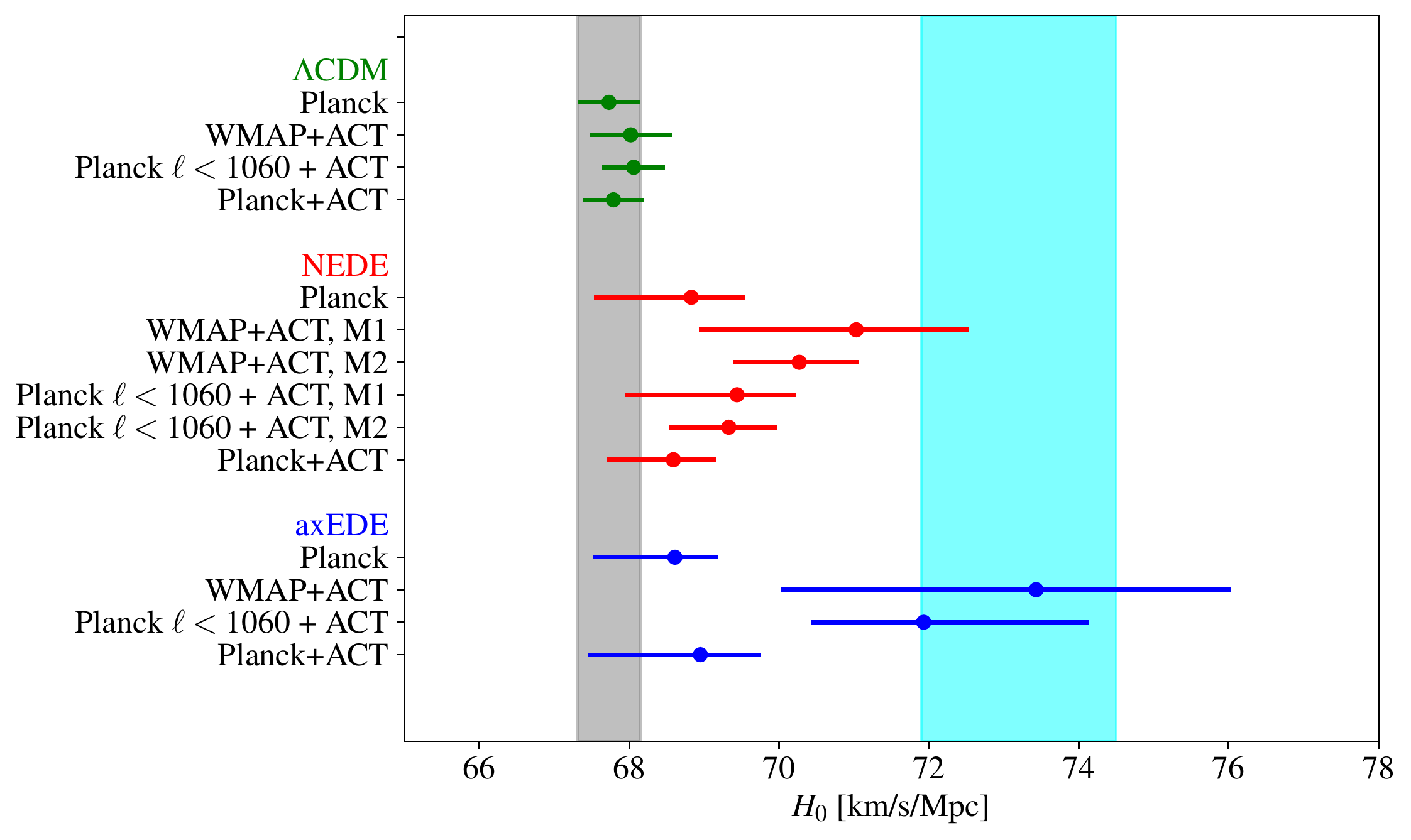}
    \caption{{\em Summary Plot.} Whisker plot at 68\% C.L. of the Hubble constant $H_0$ reconstructed in various models for various data combination, compared to the SH0ES measurements (cyan band) and the baseline $\Lambda$CDM prediction from {\it Planck} data (gray band). None of the data combinations include a late-time prior on $H_0$, but they do include BAO and Pantheon. The notation `M1' and `M2' refers to a low- and high-trigger mass mode in the NEDE model.}
    \label{fig:intro}
\end{figure*}

Our main results are summarized in Fig.~\ref{fig:intro}, where we show the posterior distribution for the Hubble constant $H_0$ using $\Lambda$CDM, axEDE and NEDE against the various data combinations considered in this work and \emph{without including a late-time prior on $H_0$}. One can clearly see a difference between the fiducial {\it Planck} analysis (favoring low-$H_0$), and the analysis of ACT data with either WMAP or restricted {\it Planck} data (favoring high-$H_0$). This illustrates the impact of {\it Planck} high-$\ell$ TT data on such model.

The paper is structured as follows. In Sec.~\ref{sec:II}, we introduce the EDE models, describe the data we use, and present the `WMAP+ACT' take on EDE cosmologies, compared with the standard {\it Planck} analysis. We also present the standard `{\it Planck}+ACT' analysis. We quantify the tension with SH0ES in Sec.~\ref{sec:H0-prior}, and compare the cosmologies obtained when including a SH0ES prior. In Sec.~\ref{sec:Planck-vs-ACT} we explore `what is it about {\it Planck} that disfavors EDE?', performing various analyses with restricted {\it Planck} data. In Sec.~\ref{sec:mock}, focusing on axEDE, we use a mock ACT data analysis to firmly establish that axEDE is not artificially favored over \LCDM. We present our conclusions in Sec.~\ref{sec:concl}. In Appendix \ref{ap:full} we demonstrate that the use of the `lite' version of the ACT likelihood is justified when exploring constraints to axEDE, and by extension other EDE cosmologies. In Appendix \ref{app:resPlanck}, we present extended results comparing the WMAP+ACT analysis to those using a restricted {\it Planck} data set.

\section{The `WMAP+ACT' take on EDE cosmologies}
\label{sec:II}

\subsection{The models}

In order to test the robustness of our conclusions to the detailed EDE dynamics, we make use of two different models, which both introduce three free parameters in addition to the standard six in $\Lambda$CDM. One is the phenomenological axion potential introduced in Refs.~\cite{Kamionkowski:2014zda,Karwal:2016vyq,Poulin:2018dzj,Poulin:2018cxd,Smith:2019ihp}, which we will refer to as `axEDE', and the other is a model of a scalar field phase transition, `new early dark energy' (NEDE) \cite{Niedermann:2019olb,Niedermann:2020dwg}. Both models are able to raise the value of $H_0$ inferred from CMB+BAO+SN1a data sets when a prior on $H_0$ is included but have different detailed dynamics. We note that these two models are far from exhaustive. For example, there are EDE models which involve non-minimal coupling to gravity \cite{Sakstein:2019fmf,Braglia:2020bym,Karwal:2021vpk}, new interactions with neutrinos \cite{Gogoi:2020qif} or a non-minimal kinetic term\footnote{For earlier investigations of an EDE cosmology, see also Refs.~\cite{Doran:2000jt,Wetterich:2004pv,Verde:2016wmz}.} \cite{Lin:2019qug,Lin:2020jcb} (and see Ref.~\cite{Schoneberg:2021qvd} for an exhaustive list). Given our conclusions, it would be interesting to fit these models to ACT data as well.

The axEDE model proposes the existence of a cosmological scalar field with a potential of the form
\begin{equation}\label{eq:potential}
    V(\theta) = m^2 f^2[1-\cos (\theta)]^3,
\end{equation} where $m$ represents the axion mass, $f$ the axion decay constant, and $\theta \equiv \phi/f$ is a re-normalized field variable defined such that $-\pi \leq \theta \leq \pi$. 
A more detailed description of this model can be found in Refs.~\cite{Poulin:2018cxd,Smith:2019ihp}, and we make use of the modified \texttt{CLASS} \cite{Blas:2011rf} publicly available at \url{https://github.com/PoulinV/AxiCLASS}. 
 As done in past work, we trade the `theory parameters', $m$ and $f$, for `phenomenological parameters', namely the critical redshift at which the field becomes dynamical $z_c$, the fractional energy density contributed by the field at the critical scale factor $f_{\rm axEDE}(z_c)$. 
 Our third parameter is $\theta_i$, which controls the effective sound speed $c_s^2$ and thus the dynamics of perturbations (mostly).  Moreover, we assume that the field always starts in slow-roll (as enforced by the very high value of the Hubble rate at early times), and without loss of generality we restrict $0\leq \theta_i \leq \pi$. 
 
The NEDE model proposes two cosmological scalar fields, the NEDE field $\psi$ of mass $M$ and the `trigger' (sub-dominant) field $\phi$ of mass $m$, whose potential is written as (with canonically normalized kinetic terms):
\begin{equation}
    V(\psi,\phi)=\frac{\lambda}{4}\psi^4+\frac{1}{2}\beta M^2\psi^2-\frac{1}{3}\alpha M\psi^3+\frac{1}{2}m^2\phi^2+\frac{1}{2}\gamma\phi^2\psi^2.
\end{equation}
where $\lambda$, $\beta$, $\alpha$, $\gamma$ are dimensionless couplings. When $H\lesssim m$, $\phi$ rolls down the potential eventually dropping below a threshold value for which the field configuration with $\psi=0$ becomes unstable, at which point a quantum tunneling to a true vacuum occurs and the energy density contained in the NEDE field rapidly dilutes.
We make use of the modified \texttt{CLASS} version presented in Ref.~\cite{Niedermann:2020dwg} and available at \url{https://github.com/flo1984/TriggerCLASS}. The NEDE model is specified by the fraction of NEDE before the decay, $f_{\rm NEDE}(z_*)\equiv\bar\rho_{\rm NEDE}(z_*)/\bar\rho_{\rm tot}(z_*)$ (where $z_*$ is given by the redshift at which $H=0.2 m$), the mass of the trigger field\footnote{In the following, we will use the simpler notation $m_{\rm NEDE}\equiv m_\phi\times$Mpc. } $m_\phi$ which controls the redshift of the decay $z_*$, and the equation of state after the decay $w_{\rm NEDE}$. We follow Ref.~\cite{Niedermann:2020dwg} and take the effective sound speed in the NEDE fluid $c_s^2$ to be equal to the equation of state after the decay, i.e. $c_s^2=w_{\rm NEDE}$. In order to be more generic and to compare two EDE models with the same number of free parameters, in our main analysis we take $w_{\rm NEDE}$ to be a free parameter with a range between $1/3$ and $1$. In Appendix \ref{app:NEDE_w23}, we also explore $w_{\rm NEDE} = 2/3$, as is done in Ref.~\cite{Niedermann:2020dwg}.

\subsection{Details of the MCMC analyses}

Our baseline analysis includes the TT, TE, and EE power spectra from ACT DR4 \cite{ACT:2020frw} and the WMAP 9 year data release \cite{WMAP:2012fli}. Following the ACT collaboration, we leave out the low-$\ell$ EE power spectra and instead place a Gaussian prior on the optical depth\footnote{This differs from the ACT collaboration's choice of $\tau =0.065\pm0.015$ \cite{Aiola:2020azj}, but is in agreement with that of the SPT collaboration \cite{SPT:2021slg}. Our conclusions are not affected by the choice of prior on $\tau$.} $\tau =0.0543\pm0.0073$, as derived from Planck data within $\Lambda$CDM \cite{Planck:2019nip}. 
Additionally, our baseline analysis includes BAO measurements from 6dFGS at $z=0.106$~\cite{Beutler:2011hx}, SDSS DR7 at $z=0.15$~\cite{Ross:2014qpa} and BOSS DR12 at $z=0.38, 0.51, 0.61$~\cite{Alam:2016hwk}, as well as un-calibrated luminosity distance of SNIa in the Pantheon catalog, spanning redshifts $0.01 < z < 2.3$~\cite{Scolnic:2017caz}. 

\begin{figure*}[ht]
    \centering
\includegraphics[width=\columnwidth]{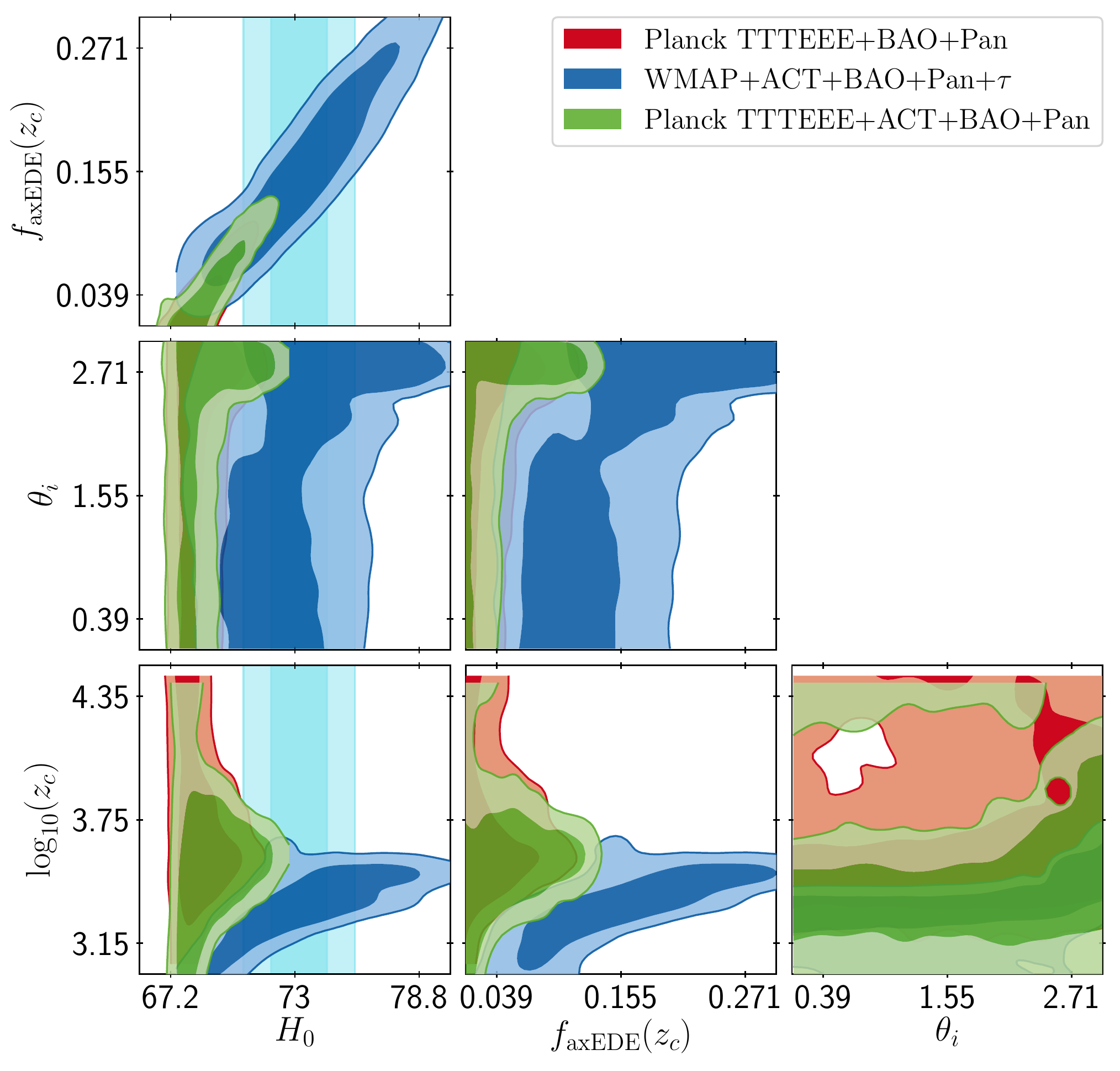}
\includegraphics[width=\columnwidth]{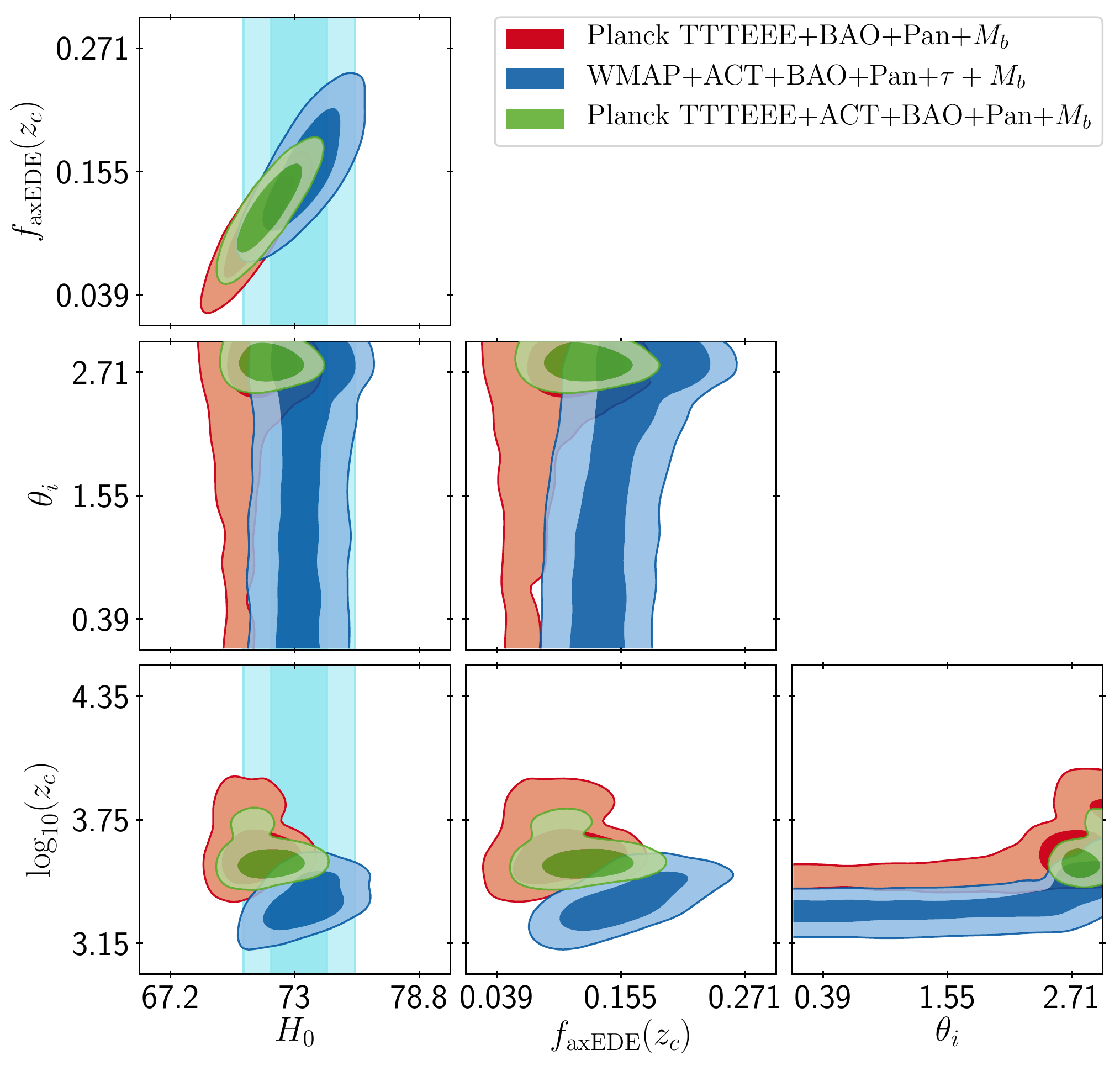}
    \caption{2D posteriors of a subset of parameters in the axEDE cosmology fit to various data sets (see legend) with (right panel) and without (left panel) a prior taken from SH0ES on the intrinsic SN1a magnitude.}
    \label{fig:MCMC_axEDE}
\end{figure*}

We also considered the impact of using different combinations of CMB data sets, leaving the BAO and SN1a data fixed. To compare to the current literature, we used {\it Planck} low-$\ell$ CMB TT, EE, the high-$\ell$ TT, TE, EE and lensing data \cite{Planck:2019nip} (including BAO and Pantheon data). We also used a combination of {\it Planck} and (restricted) ACT data. In that case, to limit double counting of information, we follow the procedure of the ACT collaboration and truncate multipoles $\ell < 1800$ in the ACT TT data\footnote{We note that, as discussed in Ref.~\cite{Aiola:2020azj}, when combining Planck and ACT we remove the `low'-$\ell$ ACT TT bins. This procedure was developed with respect to $\Lambda$CDM, and it is possible that when fitting a different cosmological model this procedure includes unaccounted for correlations between these data sets. Since our main conclusions do not rely on this combination of data we do not explore this further.}. Finally, we consider a restricted {\it Planck} + ACT data set, where the {\it Planck} multipoles are limited to $\ell \leq 1060$ (mimicking the multipole range of WMAP). 
For both axEDE and NEDE, we adopt uninformative flat priors on $\Lambda$CDM parameters, and set two massless and one massive active neutrino species with $m_{\nu} = 0.06 \ \text{eV}$, following {\emph{Planck}}'s conventions~\cite{Planck:2019nip}. We also model non-linear corrections in the matter power spectrum through the Halofit algorithm \cite{Takahashi:2012em,Ali-Haimoud:2012fzp}, as implemented in \texttt{CLASS}.
In the axEDE case, we take the priors $f_{\rm axEDE}(z_c)\in[0,0.3]$, $\log_{10}(z_c)\in[2,4.5]$ and $\theta_i \in [0,3.1]$, while in the NEDE we take the priors $f_{\rm NEDE}(z_*)\in[0,0.3]$, ${\rm log}_{10}(m_{\rm NEDE})=[1.3,3.3]$ and $w_{\rm NEDE}\in[1/3,1]$. 
We take our MCMC chains to be converged when the Gelman-Rubin criterion\footnote{Most of our chains actually have $R-1 \lesssim 0.03$, but some runs show complicated posteriors (including bi-modality), which renders a tighter $R-1$ criterion not practically achievable.} $R-1 \lesssim 0.1$ \cite{Gelman:1992zz}. 
To extract the best-fit parameters, we make use of the {\sc Minuit} algorithm \cite{James:1975dr} through the {\sc iMinuit} python package\footnote{\url{https://iminuit.readthedocs.io/}}.

\subsection{Results of the analyses}

\begin{figure*}
    \centering
\includegraphics[width=\columnwidth]{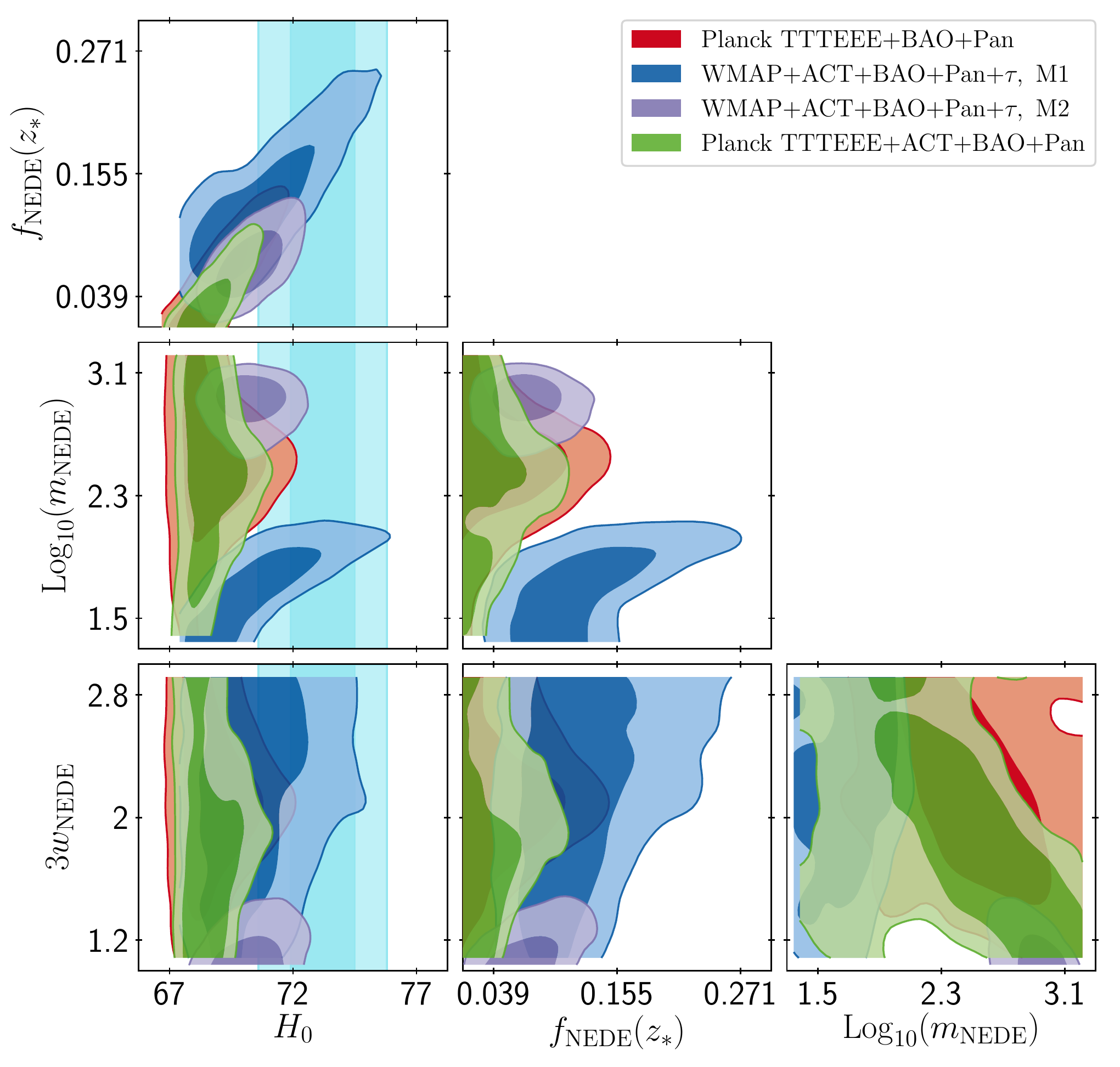}
\includegraphics[width=\columnwidth]{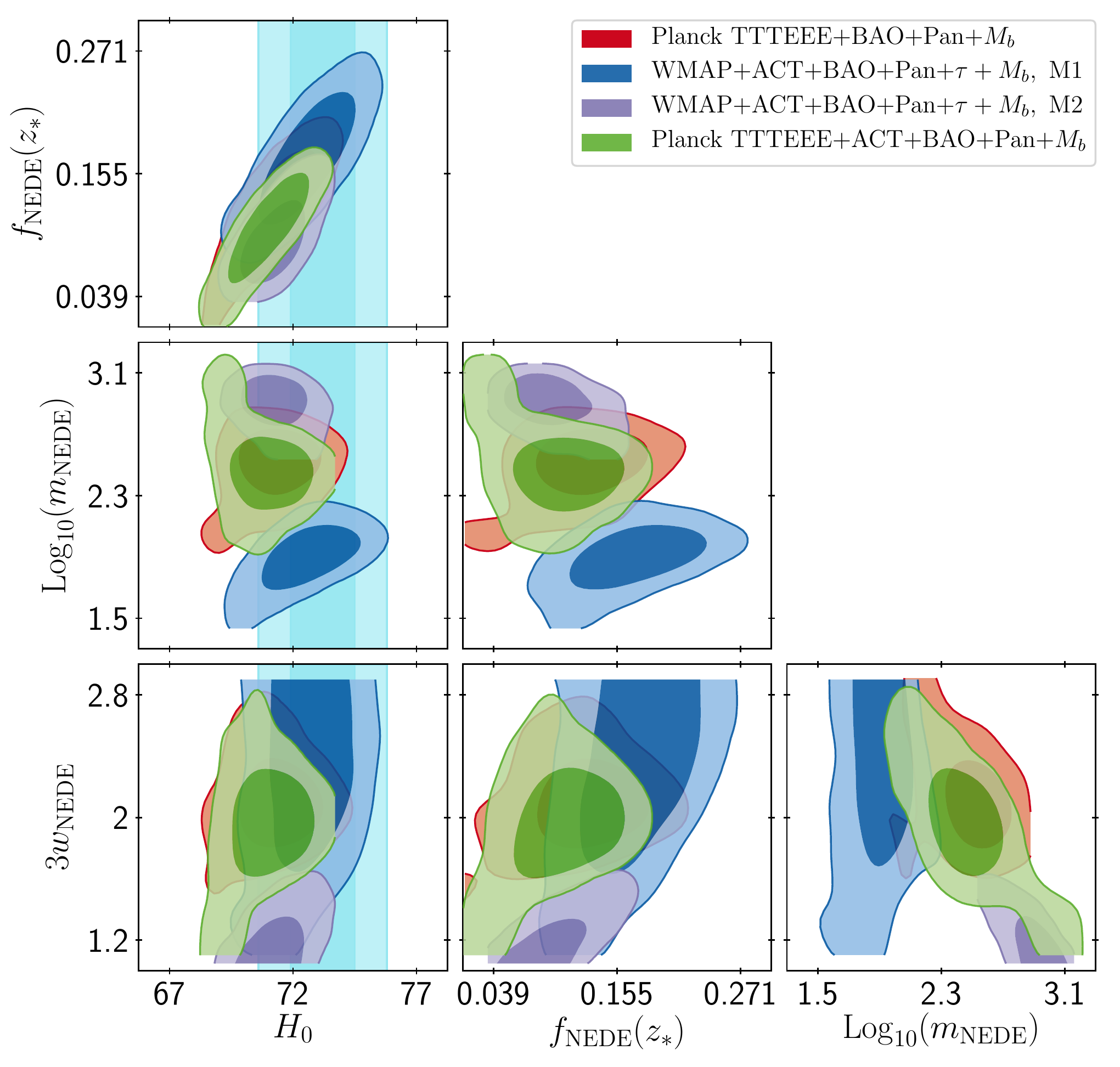}

    \caption{2D posteriors of a subset of parameters in the NEDE cosmology fit to various data sets (see legend) with (right panel) and without (left panel) a prior taken from SH0ES on the intrinsic SN1a magnitude.}
    \label{fig:MCMC_NEDE}
\end{figure*}

\begin{figure*}
    \centering
\includegraphics[width=1.5\columnwidth]{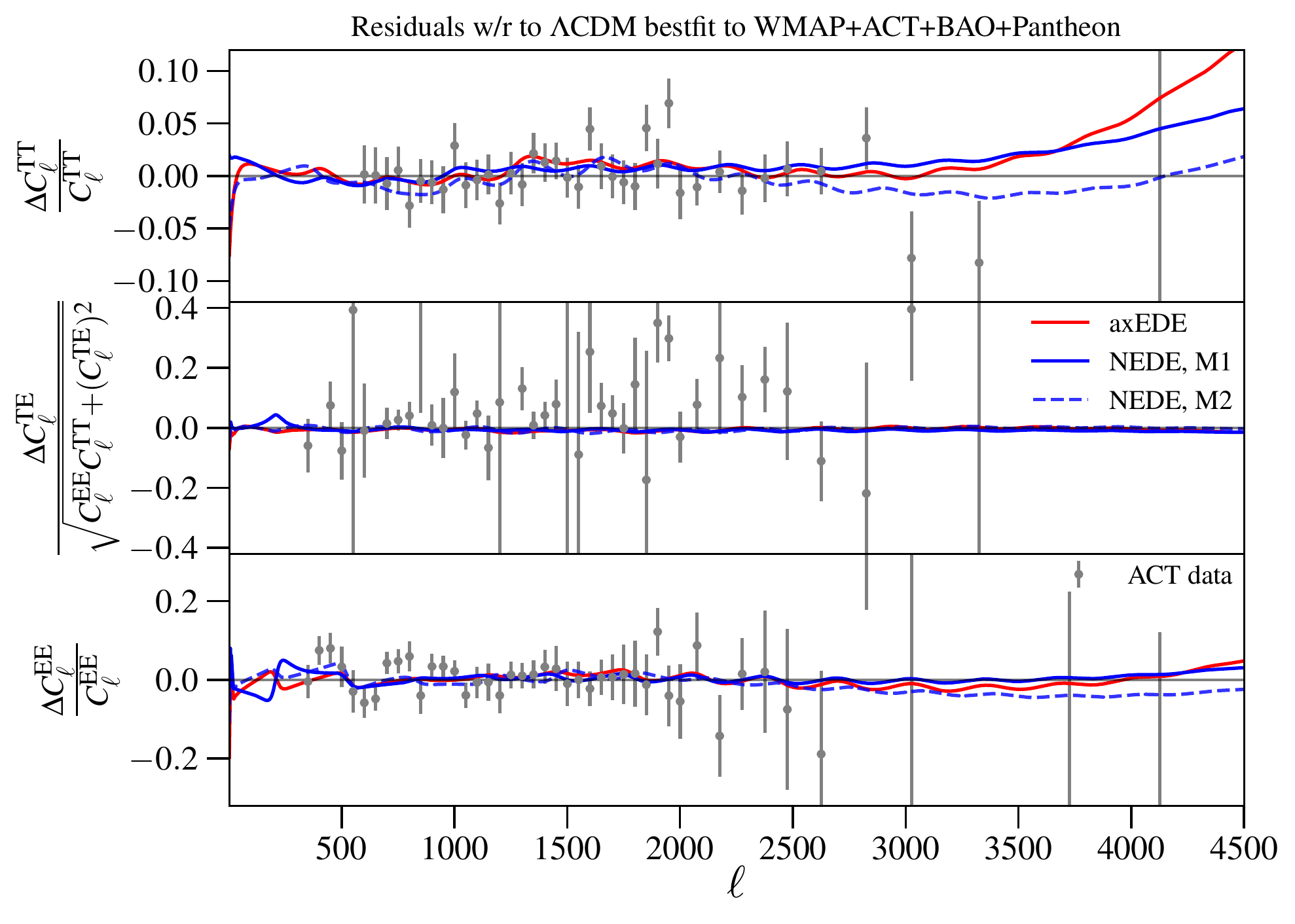}
    \caption{Residuals of the TT,TE and EE power spectra computed between the EDE cosmologies best-fit to WMAP+ACT+BAO+Pantheon and the $\Lambda$CDM best-fit to the same data combination, as reference. The data points are the residuals of ACT data with respect to the same reference $\Lambda$CDM best-fit model.}
    \label{fig:axEDE-NEDE-ACT}
\end{figure*}

We show the posterior distributions for axEDE and NEDE in Fig.~\ref{fig:MCMC_axEDE} (left panel) and \ref{fig:MCMC_NEDE} (left panel), respectively. We report all relevant results for the axEDE model in Tables \ref{tab:axEDE_WMAP} and \ref{tab:axEDE_Planck}, while results for the NEDE model are reported in Tables \ref{tab:NEDE_WMAP} and \ref{tab:NEDE_Planck}.
We report $\chi^2_{\rm min}$ per experiment for the various analyses performed in this work in App.~\ref{app:chi2}.
\begin{table*}
 \scalebox{0.95}{
 \begin{tabular}{|l|c|c|} 
 \hline
Model &  \multicolumn{2}{c|}{ Axion-like Early Dark Energy} \\
    \hline Parameter &WMAP+ACT&+SH0ES\\ 
    & +BAO+Pantheon &  \\ \hline \hline
        $f_{\rm axEDE}(z_c)$ & $0.158(0.234)_{-0.094}^{+0.051}$& $0.155(0.188)_{-0.041}^{+0.033}$  \\
    $\log_{10}(z_c)$ &  $3.326(3.493)_{-0.093}^{+0.2}$ &$3.352(3.444)_{-0.075}^{+0.1}$ \\
    $\theta_i$ & unconstrained (2.813)& unconstrained (2.815) \\
    \hline
    $H_0$ [km/s/Mpc] & $73.43(75.52)_{-3.4}^{+2.6}$&$73.44(73.94)\pm1.2$\\
    $100~\omega_b$ &$2.201(2.216)_{-0.035}^{+0.049}$ &$2.205(2.214)_{-0.033}^{+0.039}$\\
    $\omega_{\rm cdm}$& $0.1402(0.1505)_{-0.015}^{+0.0098}$& $0.1401(0.1420)_{-0.006}^{+0.0058}$\\
    $10^9A_s$ &$2.171(2.206)_{-0.058}^{+0.072}$ & $2.176(2.190)_{-0.039}^{+0.045}$ \\
    $n_s$& $0.9884(1.001)\pm0.019$ & $0.9886(0.9953)\pm0.012$ \\
    $\tau_{\rm reio}$& $0.0539(0.5269)_{-0.0068}^{+0.0074}$& $0.0535(0.0546)_{-0.0071}^{+0.0069}$ \\

    \hline
    $M_b$ & $-19.25(-19.18)_{-0.095}^{+0.082}$& $-19.24(-19.23)_{-0.035}^{+0.037}$ \\
    $S_8$ &$0.862(0.888)_{-0.034}^{+0.042}$ & $0.866(0.867)_{-0.024}^{+0.026}$\\
    $\Omega_m$ &$0.3019(0.3039)_{-0.0081}^{+0.0073}$  & $0.3018(0.3013)_{-0.0073}^{+0.0069}$\\
    \hline
    $\chi^2_{\rm min}$ &6931.7 & 6932.5\\
        $\Delta\chi^2_{\rm min}(\Lambda{\rm CDM})$ &-14.6& -31.5\\

        \hline

    $Q_{\rm DMAP}$ &  \multicolumn{2}{c|}{ $0.9\sigma$}\\
    \hline
\end{tabular} }
\caption{The mean (best-fit) $\pm 1\sigma$ errors of the cosmological parameters reconstructed from analyses of WMAP and ACT data (together with BAO and SN1a data) in the axEDE model. For each data set, we also report the best-fit $\chi^2$ and the $Q_{\rm DMAP}$ tension with SH0ES.}
\label{tab:axEDE_WMAP}
\end{table*}

\begin{table*}
 \scalebox{0.95}{
 \begin{tabular}{|l|c|c|c|c|} 
 \hline
Model &  \multicolumn{4}{c|}{ Axion-like Early Dark Energy} \\
    \hline Parameter &Planck&+SH0ES&Planck+ACT&+SH0ES\\ 
    & +BAO+Pantheon & & +BAO+Pantheon & \\ \hline \hline
        $f_{\rm axEDE}(z_c)$  & $<0.084(0.09)$&$0.103(0.125)_{-0.028}^{+0.035}$ & $< 0.11(0.118)$&$0.121(0.124)_{-0.026}^{+0.028}$ \\
    $\log_{10}(z_c)$  & unconstrained (3.569)&$3.602(3.574)_{-0.044}^{+0.11}$  &$3.417(3.498)_{-0.43}^{+0.2}$  & $3.548(3.566)_{-0.031}^{+0.049}$ \\
    $\theta_i$ & $1.933(2.773)_{-0.44}^{+1.2}$& $2.578(2.744)_{0.011}^{+0.35}$&unconstrained (2.688) &$2.794(2.803)_{-0.078}^{+0.087}$\\
    \hline
    $H_0$ [km/s/Mpc]  &$68.6(70.88)_{-1.1}^{+0.55}$  & $71.23(72.03)\pm1.1$& $68.95(71.54)_{-1.6}^{+0.76}$&  $71.79(72.16)\pm0.99$  \\
    $100~\omega_b$  &$2.257(2.270)_{-0.02}^{+0.017}$ & $2.281(2.284)_{-0.023}^{+0.02}$&$2.246( 2.261)\pm0.016$ & $2.258(2.270)\pm0.018$\\
    $\omega_{\rm cdm}$& $0.1219(0.1278)_{-0.0034}^{+0.0013}$ &$0.1297(0.1321)\pm0.0039$ & $0.1233( 0.1318)_{-0.005}^{+0.0021}$& $0.1319(0.1318)_{-0.0038}^{+0.0035}$\\
    $10^9A_s$ & $2.118(2.159)_{-0.034}^{+0.031}$& $2.149(2.159)_{-0.035}^{+0.032}$&$2.134( 2.142)_{-0.033}^{+0.031}$ & $2.161(2.168)_{-0.034}^{+0.03}$\\
    $n_s$ & $0.9719( 0.9850)_{-0.0076}^{+0.0048}$& $0.9871(0.9912)_{-0.0068}^{+0.0072}$&$0.9755(0.9873)_{-0.0088}^{+0.0058}$ & $0.9889(0.9937)_{-0.0064}^{+0.0057}$\\
    $\tau_{\rm reio}$ & $0.0569(0.0617)_{-0.0078}^{+0.0071}$&$0.05769(0.05768)_{-0.0079}^{+0.0073}$  & $0.0557(0.0517)_{-0.0077}^{+0.0068}$ &  $0.0552(0.0573)_{-0.0079}^{+0.0071}$ \\
  
    \hline
    $M_b$ &  $-19.39(-19.32)_{-0.033}^{+0.016}$ & $-19.31( -19.29)\pm0.032$ &$-19.38(-19.30)_{-0.047}^{+0.023}$ &$-19.29(-19.28)\pm0.029$  \\
    $S_8$ & $0.828(0.836)\pm0.013$& $0.839(0.843)\pm0.013$&$0.834(0.843)\pm0.013$ & $0.844(0.843)\pm0.013$\\
    $\Omega_m$ & $0.3085(0.3008)\pm0.0059$  & $0.3019(0.3000)\pm0.0056$ & $0.3079(0.3030)_{-0.0062}^{+0.0059}$  &  $0.3010(0.2980)_{-0.0053}^{+0.005}$ \\
    \hline
    $\chi^2_{\rm min}$  &3804.0 &  3806.3&4046.4 & 4046.5\\
        $\Delta\chi^2_{\rm min}(\Lambda{\rm CDM})$  &-4.0&  -21.8&-3.8 &-24.0 \\

        \hline

    $Q_{\rm DMAP}$ &   \multicolumn{2}{c|}{ $1.5\sigma$} &  \multicolumn{2}{c|}{ $0.3\sigma$} \\
    \hline
\end{tabular} }
\caption{The mean (best-fit) $\pm 1\sigma$ errors of the cosmological parameters reconstructed from analyses of {\it Planck} and ACT data (together with BAO and SN1a data) in the axEDE model. For each data set, we also report the best-fit $\chi^2$ and the $Q_{\rm DMAP}$ tension with SH0ES.}
\label{tab:axEDE_Planck}
\end{table*}

\begin{table*}
 \scalebox{0.95}{
 \begin{tabular}{|l|c|c|c|c|} 
 \hline
 Model &  \multicolumn{4}{c|}{New Early Dark Energy} \\

    \hline Parameter &WMAP+ACT&+SH0ES&WMAP+ACT&+SH0ES\\ 
    & +BAO+Pantheon (M1) &  &+BAO+Pantheon (M2)& \\
    \hline \hline
        $f_{\rm NEDE}(z_*)$ & $0.12(0.151)_{-0.055}^{+0.03}$&$0.169(0.189)_{-0.04}^{+0.043}$  & $0.071(0.065)_{-0.024}^{+0.02}$&  $0.094(0.078)_{-0.033}^{+0.019}$\\
    ${\rm log}_{10}(m_{\rm NEDE})$ &$1.687(1.850)_{-0.25}^{+0.22}$  &$1.901(1.94)_{-0.12}^{+0.15}$ &$2.916(3.009)_{-0.079}^{+0.13}$  & $2.897(3.006)_{-0.076}^{+0.15}$ \\
    $3w_{\rm NEDE}$ & $2.14(2.73)_{-0.27}^{+0.86}$& $2.28(2.76)_{-0.2}^{+0.72}$ &$1.12(1.00)_{-0.12}^{+0.02}$ & $ 1.181(1.054)_{-0.18}^{+0.042}$\\
    \hline
    $H_0$ [km/s/Mpc] & $70.51(71.57)_{-2.2}^{+1.1}$&  $72.54(72.96)_{-1.2}^{+1.3}$ & $70.3(70.32)_{-0.95}^{+0.89}$& $71.32(70.77)_{-0.96}^{+0.81}$  \\
    $100~\omega_b$ &  $2.167(2.188)_{-0.043}^{+0.04}$ &$2.189(2.206)_{-0.037}^{+0.042}$ & $2.201(2.178)_{-0.03}^{+0.026}$ & $2.215(2.193)_{-0.03}^{+0.026}$\\
    $\omega_{\rm cdm}$& $0.135(0.133)_{-0.0056}^{+0.006}$&$0.1367( 0.1366)_{-0.0057}^{+0.0051}$ &$0.1296(0.1291)_{-0.0034}^{+0.003}$ & $0.1315(0.1307)_{-0.0041}^{+0.0031}$ \\
    ${\rm ln}(10^{10}A_s)$ & $3.047(3.068)_{-0.025}^{+0.023}$ & $3.066(3.077)_{-0.019}^{+0.023}$ &$3.073(3.077)\pm0.016$ &  $3.076(3.076)\pm0.016$\\
    $n_s$& $0.9758(0.9751)_{-0.011}^{+0.013}$& $0.9903(0.9825)\pm0.0066$ & $0.9904(0.9911)_{-0.0063}^{+0.0062}$&   $0.9959(0.9941)_{-0.0069}^{+0.0065}$ \\
    $\tau_{\rm reio}$& $0.0541(0.0535)_{-0.0072}^{+0.0066}$ & $0.0534(0.0532)_{-0.007}^{+0.0065}$&$0.0547(0.0540)_{-0.0071}^{+0.0066}$ &  $0.0558(0.0528)_{-0.0076}^{+0.0066}$  \\

    \hline
        $M_b$ &$-19.33(-19.30)_{-0.067}^{+0.034}$  & $-19.27(-19.26)_{-0.036}^{+0.041}$&$-19.34(-19.34)_{-0.027}^{+0.026}$ & $-19.31(-19.32)_{-0.029}^{+0.023}$\\

    $S_8$ & $0.816(0.849)_{-0.037}^{+0.027}$& $0.845(0.858)_{-0.02}^{+0.03}$ &$0.842(0.849)_{-0.012}^{+0.012}\pm0.019$ & $0.847(0.849)_{-0.014}^{+0.012}$ \\
    $\Omega_m$ & $0.3031(0.3015)_{-0.0074}^{+0.007}$&$0.2993(0.2980)_{-0.0076}^{+0.0072}$ &  $0.3074(0.3051
    )_{-0.008}^{+0.0074}$&  $0.3034(0.3032)_{-0.0076}^{+0.0072}$\\
    $z_*$ & $1573(1872)_{-680}^{+210}$ & $2023(2066)_{-380}^{+340}$  & $7870(8774)_{-900}^{+1200}$ & $7642(8617)_{-890}^{+1400}$\\
    \hline
    $\chi^2_{\rm min}$ &6937.9&  6938.9& 6928.7&  6934.5\\
            $\Delta\chi^2_{\rm min}(\Lambda{\rm CDM})$  &-8.4  &  -25.1&-17.6 & -29.5\\

     \hline
    $Q_{\rm DMAP}$ & \multicolumn{2}{c|}{ $1.0\sigma$} & \multicolumn{2}{c|}{ $2.4\sigma$} \\
    \hline
\end{tabular} }
\caption{The mean (best-fit) $\pm 1\sigma$ errors of the cosmological parameters reconstructed  from analyses of WMAP and ACT data (together with BAO and SN1a data) in the NEDE model. For each data set, we also report the best-fit $\chi^2$ and the $Q_{\rm DMAP}$ tension with SH0ES. We recall that $m_{\rm NEDE}\equiv m_\phi\times$Mpc.}
\label{tab:NEDE_WMAP}
\end{table*}

\begin{table*}
 \scalebox{0.95}{
 \begin{tabular}{|l|c|c|c|c|} 
 \hline
  Model &  \multicolumn{4}{c|}{New Early Dark Energy} \\

    \hline Parameter &Planck&+SH0ES&Planck+ACT&+SH0ES\\ 
    & +BAO+Pantheon & & +BAO+Pantheon &  \\
    \hline \hline
        $f_{\rm NEDE}(z_*)$ & $<0.116(0.042)$ & $0.125(0.138)_{-0.028}^{+0.037}$& $<0.086(0.027)$ &$0.102(0.107)_{-0.029}^{+0.04}$ \\
    ${\rm log}_{10}(m_{\rm NEDE})$   &$2.52(2.50)\pm0.34$ & $2.5(2.622)_{-0.089}^{+0.17}$ &  unconstrained (3.167)& $2.47(2.43)_{-0.18}^{+0.15}$\\
    $3w_{\rm NEDE}$  & $2.11(1.84)_{-0.44}^{+0.5}$&  $2.11(1.96)_{-0.2}^{+0.14}$& unconstrained (1.089) & $1.96(1.99)_{-0.23}^{+0.25}$\\
    \hline
    $H_0$ [km/s/Mpc]&$68.83(69.05)_{-1.3}^{+0.69}$ &  $71.32(71.66)_{-0.96}^{+1.1}$&  $68.62(68.89)_{-0.98}^{+0.59}$&  $70.94(70.94)_{-1}^{+1.2}$  \\
    $100~\omega_b$ &$2.258(2.259)_{-0.026}^{+0.019}$ & $2.293(2.303)_{-0.024}^{+0.026}$ & $2.242( 2.247)_{-0.018}^{+0.016}$&  $2.267(2.264)\pm0.02$ \\
    $\omega_{\rm cdm}$&$0.1232(0.1226)_{-0.004}^{+0.0019}$ &  $0.1299(0.1315)_{-0.0034}^{+0.0037}$&$0.1221(0.1230)_{-0.003}^{+0.0017}$ & $0.1279(0.1283)_{-0.0032}^{+0.0038}$\\
    ${\rm ln}(10^{10}A_s)$  & $3.056(3.055)_{-0.016}^{+0.014}$& $3.07(3.08)_{-0.016}^{+0.015}$&$3.057(3.071)_{-0.016}^{+0.013}$ &$3.069(3.075)_{-0.015}^{+0.014}$   \\
    $n_s$& $0.9734(0.9751)_{-0.0087}^{+0.0058}$ & $0.9884(0.9920)_{-0.0065}^{+0.0077}$& $0.9741(0.9824)\pm0.0066$&  $0.9871(0.9888)_{-0.0073}^{+0.0071}$ \\
    $\tau_{\rm reio}$ & $0.0572(0.0565)_{-0.0076}^{+0.0067}$&  $0.0582(0.0634)_{-0.008}^{+0.0067}$&$0.0550(0.0608)_{-0.0074}^{+0.0066}$ &  $0.05575(0.0582)_{-0.0074}^{+0.0062}$ \\

    \hline
    $M_b$ & $-19.38(-19.38)_{-0.04}^{+0.02}$& $-19.31(-19.30)\pm0.03$&$-19.39(-19.38)_{-0.028}^{+0.017}$ & $-19.32(-19.32)_{-0.031}^{+0.035}$\\
    $S_8$  & $0.833(0.827)_{-0.013}^{+0.011}$& $0.842(0.850)\pm0.013$ &  $0.836(0.842)\pm0.012$&$0.837(0.845)_{-0.011}^{+0.012}$ \\
    $\Omega_m$ & $0.309(0.3060)\pm0.006$& $0.3017(0.3023)\pm0.0055$& $0.3097(0.3065)_{-0.0056}^{+0.0058}$ & $0.3004(0.3000)_{-0.006}^{+0.0055}$ \\
    $z_*$ &  $5238(4632)_{-2800}^{+1200}$ &$4591(5257)_{-660}^{+880}$  &unconstrained(10808.72) &  $4489(4152)_{-1200}^{+730}$\\
    \hline
    $\chi^2_{\rm min}$  & 3806.3& 3809.3& 4043.8& 4052.4\\
            $\Delta\chi^2_{\rm min}(\Lambda{\rm CDM})$  &-1.8 &  -18.8& -6.3 &-18.1\\

     \hline
    $Q_{\rm DMAP}$  & \multicolumn{2}{c|}{ $1.7\sigma$} & \multicolumn{2}{c|}{ $2.9\sigma$} \\
    \hline
\end{tabular} }
\caption{The mean (best-fit) $\pm 1\sigma$ errors of the cosmological parameters reconstructed from analyses of {\it Planck} and ACT data (together with BAO and SN1a data) in the NEDE model. For each data set, we also report the best-fit $\chi^2$ and the $Q_{\rm DMAP}$ tension with SH0ES. We recall that $m_{\rm NEDE}\equiv m_\phi\times$Mpc.}
\label{tab:NEDE_Planck}
\end{table*}

First, one can see that the results of the WMAP+ACT analysis are significantly different from 
those of the {\it Planck} analysis. 
When fit to WMAP+ACT we find that a non-zero contribution of axEDE and NEDE are favored at $\gtrsim2\sigma$. We stress that this preference for a non-zero axEDE and NEDE contribution does not rely on the inclusion of a prior from SH0ES on the value of $H_0$ (since we did not include such prior).
On the other hand, {\it Planck} leads only to upper limit on $f_{\rm axEDE}(z_c) < 0.089$ and $f_{\rm NEDE}(z_*) < 0.116$ (95\% C.L.), in agreement with previous works \cite{Smith:2019ihp,Smith:2020rxx,Niedermann:2020dwg}.

\textbf{Axion-like Early Dark Energy:} In the axEDE case, we find $f_{\rm axEDE}(z_c) = 0.158_{-0.094}^{+0.051}$ and $\log_{10}(z_c)=3.326_{-0.093}^{+0.2}$, while $\theta_i$ is unconstrained. Remarkably, $H_0 = 73.43_{-3.4}^{+2.6}$ and is in  agreement with the SH0ES determination of $H_0$. Comparing with $\Lambda$CDM, we find $\Delta\chi^2_{\rm min}({\rm axEDE}) = \chi^2_{\rm min}({\Lambda}{\rm CDM})-\chi^2_{\rm min}({\rm axEDE}) = -14.6$, for 3 extra free parameters. Despite the difference between WMAP+ACT and {\it Planck}, it is instructive to replace WMAP by {\it Planck} given that the inconsistency is mild, and to attempt to further constrain the axEDE contribution. Interestingly, we find that the combination of {\it Planck}+ (TT-restricted) ACT leads to a {\em weaker} $95\%$ C.L. upper limit  ($f_{\rm axEDE}<0.110$) than without ACT ($f_{\rm axEDE}<0.084$). 

\textbf{New Early Dark Energy:} In the NEDE case, the distribution is more complicated, showing a bi-modality in $m_{\rm NEDE}$. 
To better capture the two modes, we perform two MCMC analyses, splitting the parameter space between $\log_{10}(m_{\rm NEDE})\in [ 1.3, 2.5]$ (which we denote by `M1') and
$\log_{10}(m_{\rm NEDE})\in [2.5,3.3]$ (`M2').
We find that the high-mass mode has $\log_{10}(m_{\rm NEDE})=2.933_{-0.08}^{+0.13}$ (corresponding to $z_* = 8040_{-930}^{+1200}$) with associated $f_{\rm NEDE}(z_*) = 0.071_{-0.022}^{+0.016}$, and $H_0 = 70.26_{-0.87}^{+0.77}$ km/s/Mpc, while the low-mass mode has $\log_{10}(m_{\rm NEDE}) = 1.779_{-0.16}^{+0.26}$ with $f_{\rm NEDE}(z_*) = $ and $H_0 = 71.05_{-2.2}^{+1.4}$ km/s/Mpc. 
The high-mass mode represents an improvement with respect to $\Lambda$CDM of $\Delta\chi^2_{\rm min}({\rm NEDE})=-17.6$ while the low-mass mode has $\Delta\chi^2_{\rm min}({\rm NEDE})=-8.4$. Note that the high-mass mode -- with slightly lower $H_0$ -- has a significantly lower $\chi^2$ than the low-mass mode, and is thus favored over the mode that would fully resolve the Hubble tension.
Finally, combining {\it Planck} with (TT-restricted) ACT, we find that the NEDE model still improves the fit over $\Lambda$CDM by a small amount, $\Delta\chi^2_{\rm min}(\Lambda{\rm CDM})\simeq -5.7$, but $f_{\rm NEDE}(z_*)$ is compatible with zero at $1\sigma$. In fact, the $2\sigma$ constraint on the NEDE contribution significantly strengthens, from  $f_{\rm NEDE}(z_*) < 0.116$ (without ACT) to $f_{\rm NEDE}(z_*)<0.082$ (with ACT). This is in contrast with the result for axEDE, and indicates that the combination of {\it Planck} and ACT has the potential to disentangle between different EDE cosmologies.  This is because within {\it Planck}, the low-mass mode with high-$f_{\rm NEDE}(z_*)$ and high-$H_0$ is not present. Given that the high-mass mode within ACT -- with a smaller $f_{\rm NEDE}(z_*)$ -- also has a smaller $\chi^2$, the combination of {\it Planck} and ACT favors this mode, leading to a smaller upper limit on $f_{\rm NEDE}(z_*)$. 

Nevertheless, the difference between the WMAP+ACT and {\it Planck} 2D posteriors for the EDE cosmologies hints that the two data sets may have some features which are inconsistent. In Sec.~\ref{sec:Planck-vs-ACT}, we further establish what features disfavor the WMAP+ACT EDE cosmologies within {\it Planck} data.

\subsection{What about ACT seems to favor EDE?}

It is of interest to assess what features of ACT favor EDE over $\Lambda$CDM. Fig.~\ref{fig:axEDE-NEDE-ACT} shows the residuals of the axEDE and NEDE best-fit to WMAP+ACT+BAO+Pantheon with respect to the $\Lambda$CDM best-fit to the same data combination\footnote{Our best-fit parameters are $\{100\omega_b=2.236,        \omega_{cdm}=0.1193,               H_0 = 67.87,              n_s=0.9736,              10^9 A_s=2.1108,         \tau_{\rm reio}=0.0540\}$}. The data points show the residuals of the ACT data with respect to the best-fit $\Lambda$CDM model to that same data combination. 

First, focus on the residual of the EDE models. Within the accuracy of ACT measurements, one can see that the TE power spectrum does not show any identifiable features (in fact it is remarkably consistent with zero). Therefore, the ACT TE power spectrum does not discriminate between the different models.  On the other hand both TT and EE power spectra have interesting features. In EE, all three EDE curves show noticeable deviations at $\ell \lesssim 500$, as well as at $\ell \gtrsim 2500$. In TT, the deviations oscillate around zero below $\ell\sim 2000$, and become increasingly important above $\sim 2500$. Note that the high-$\ell$ tail above $\sim 3000$ in each model is very different from one another, but the current ACT precision is not sufficient to differentiate between them. 

Next, focus on the ACT residuals. Data points which deviate from zero indicate where the \LCDM\ model deviates from the measurements and therefore the EDE model might fit these data points better. There is an oscillation in the ACT TT residuals from $500  \lesssim \ell \lesssim 2000$ which appears to track oscillations in the EDE curves. However, since these data points are all statistically consistent with zero, they are unlikely to have much statistical weight when comparing the fit to these models. In the EE spectrum there are clusters of bins between $300 \lesssim \ell \lesssim 1000$ which are systematically offset from zero. We can see that the EDE curves have a shift upwards at the lower end of this range, providing a slightly better `fit-by-eye' to the data there than \LCDM.

None of these features truly appear as `smoking gun' of an EDE given the accuracy of ACT. From these residuals, it is clear that higher accuracy measurement of the tail of the TT (and EE) power spectrum, as well as large scale EE measurement (between $\ell \sim 300-500$) will help differentiating between EDE models (and $\Lambda$CDM). 

\section{Quantifying the tension with SH0ES and the agreement between high-$H_0$ EDE cosmologies}
\label{sec:H0-prior}

Before moving on to analyzing differences between {\it Planck} and WMAP+ACT, it is interesting to compare these `high-$H_0$' cosmologies to those obtained when including information from SH0ES. We thus repeat the analysis presented above, and following Refs.~\cite{Benevento:2020fev,Camarena:2021jlr,Efstathiou:2021ocp}, we now include a prior from the SH0ES collaboration on the SN1a intrinsic magnitude $M_b$. Comparing the $\chi^2$ with and without the $M_b$ prior will allow us to compute the $Q_{\rm DMAP}$ tension metric introduced in Ref.~\cite{Raveri:2019mxg}.
In the right-hand panel of Fig.~\ref{fig:MCMC_axEDE} and \ref{fig:MCMC_NEDE} we plot the results of these analysis for the axEDE and NEDE, respectively.

\textbf{Axion-like Early Dark Energy:} In the axEDE case, we find a good agreement between all reconstructed parameters. The most notable difference is in the redshift of the transition, where {\it Planck}+SH0ES favors  $\log_{10}(z_c)=3.602_{-0.044}^{+0.11}$, while WMAP+ACT favors a slightly lower value, $\log_{10}(z_c)=3.326_{-0.093}^{+0.2}$, i.e. a transition slightly {\em after} matter-radiation equality. Yet, the combination {\it Planck}+ACT+SH0ES  improves the precision to  $\log_{10}(z_c)=3.548_{-0.031}^{+0.049}$. On the other hand, the initial field value is unconstrained with WMAP+ACT, which is different from the results with {\it Planck}, that favors high initial field values (and a peculiar shape for the potential $d^2 V/d \theta^2$ close to the initial field value) \cite{Lin:2019qug,Smith:2019ihp}. Nevertheless, the combination  {\it Planck}+ACT+SH0ES  leads to the tight measurement $\theta_i = 2.794_{-0.078}^{+0.087}$, which is remarkably consistent with the {\it Planck}-only result, while being much more constrained. This further establishes the fact that CMB data are particularly sensitive to the dynamics of EDE perturbations around the time where the field becomes dynamical \cite{Lin:2019qug,Smith:2019ihp} (in particular EE data). Finally, {\it Planck}+ACT+SH0ES leads to a $\sim 4.8\sigma$ detection of axEDE, with $f_{\rm axEDE}(z_c)= 0.12\pm0.028$, an improvement from the run without ACT, which led to $f_{\rm axEDE}(z_c) = 0.10_{-0.028}^{+0.035}$.  
Additionally, we quantify the $Q_{\rm DMAP}$-tension metric between WMAP+ACT+BAO+Pantheon and SH0ES, computed from $\sqrt{\Delta\chi^2_{\rm min}}$ between the analyses with and without the SH0ES prior, to be $\sim0.9\sigma$. This clearly shows that these data sets are in statistical agreement when fit to the axEDE cosmology, while when fit to $\Lambda$CDM, we find a $\sim4.2\sigma$ tension. Remarkably, there is also no tension between {\it Planck}+ACT+BAO+Pantheon and SH0ES, given that we find a $Q_{\rm DMAP}\simeq 0.3\sigma$. For comparison, the tension within the $\Lambda$CDM model is $4.5\sigma$. We conclude that, even when adopting a conservative analysis and combining the mildly discrepant {\it Planck} and ACT data set, the axEDE model provides an excellent resolution to the Hubble tension.

\textbf{New Early Dark Energy:} The case of NEDE is very different. First, we note that the low-mass mode M1 is in good agreement with SH0ES with $Q_{\rm DMAP}\simeq 1.0\sigma$, while the high-mass mode M2 (with the lowest $\chi^2_{\rm min}$) is still in $2.4\sigma$ tension. Moreover, as can be seen in the right-panel of Fig.~\ref{fig:MCMC_NEDE}, the region of parameter space favored by {\it Planck}+BAO+Pantheon+$M_b$ lives between the two $\log_{10}(m_{\rm NEDE})$-modes that are favored by WMAP+ACT+BAO+Pantheon. As a result, the $Q_{\rm DMAP}$ estimator indicates that the combination {\it Planck}+ACT+BAO+Pantheon is in $\sim 3\sigma$ tension with SH0ES within the NEDE model, and the reconstructed NEDE fraction drops by $\sim1\sigma$, from  $f_{\rm NEDE}(z_*)=0.129_{-0.031}^{+0.034}$ (without ACT) to $f_{\rm NEDE}(z_*)=0.108_{-0.027}^{+0.029}$ (with ACT). We conclude that in light of the combination of {\it Planck} and ACT data, the NEDE model does not provide as good a resolution to the Hubble tension. Naturally, these conclusions could be affected by the mild discrepancy between {\it Planck} and ACT we further explore in the next section.  Additionally, it is possible that different conclusions would be reached if we had followed the approach from Ref.~\cite{Niedermann:2020dwg} to fix $w_{\rm NEDE}=2/3$. In fact, we note that the analysis {\it Planck}+ACT+BAO+Pantheon+SH0ES analysis seem to favor that value, while the analysis without SH0ES favor smaller values of $w_{\rm NEDE}$, but not incompatible with $w_{\rm NEDE}=2/3$. We present an analysis of the $w_{\rm NEDE}=2/3$ case against {\it Planck}+ACT+BAO+Pantheon in App.~\ref{app:NEDE_w23}.

\begin{figure*}
    \centering
    \includegraphics[width=1.5\columnwidth]{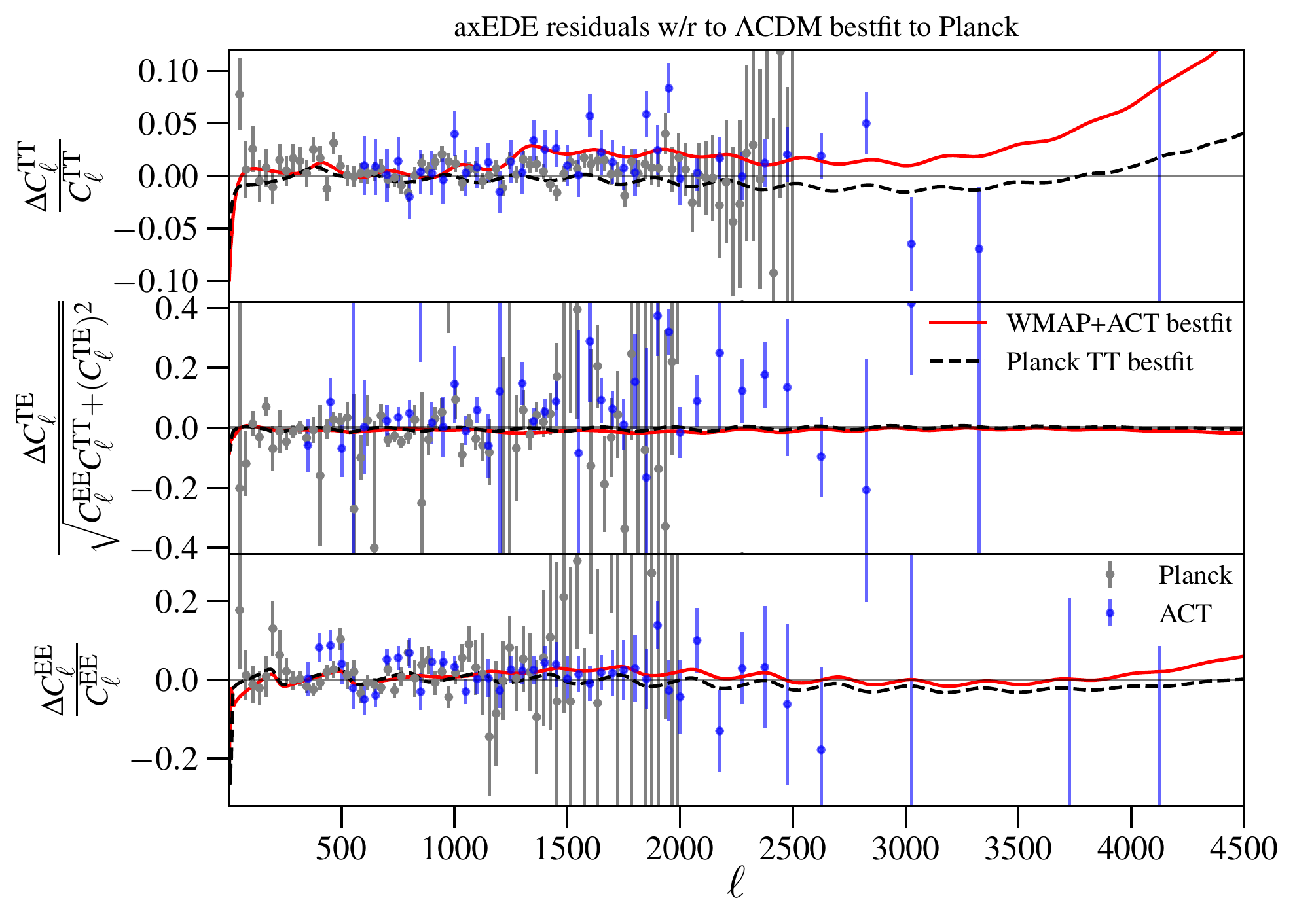}
    \caption{Residual of the CMB power spectra between $\Lambda$CDM from {\it Planck} and the axEDE best-fit cosmologies of the `WMAP+ACT' or '{\it Planck} highTT+lowEE+lowTT' data combination. We also show the {\it Planck} and ACT data residuals with respect to the {\it Planck} best-fit $\Lambda$CDM model.}
    \label{fig:axEDE_Cls}
\end{figure*}
\begin{figure*}
    \centering
            \includegraphics[width=1.5\columnwidth]{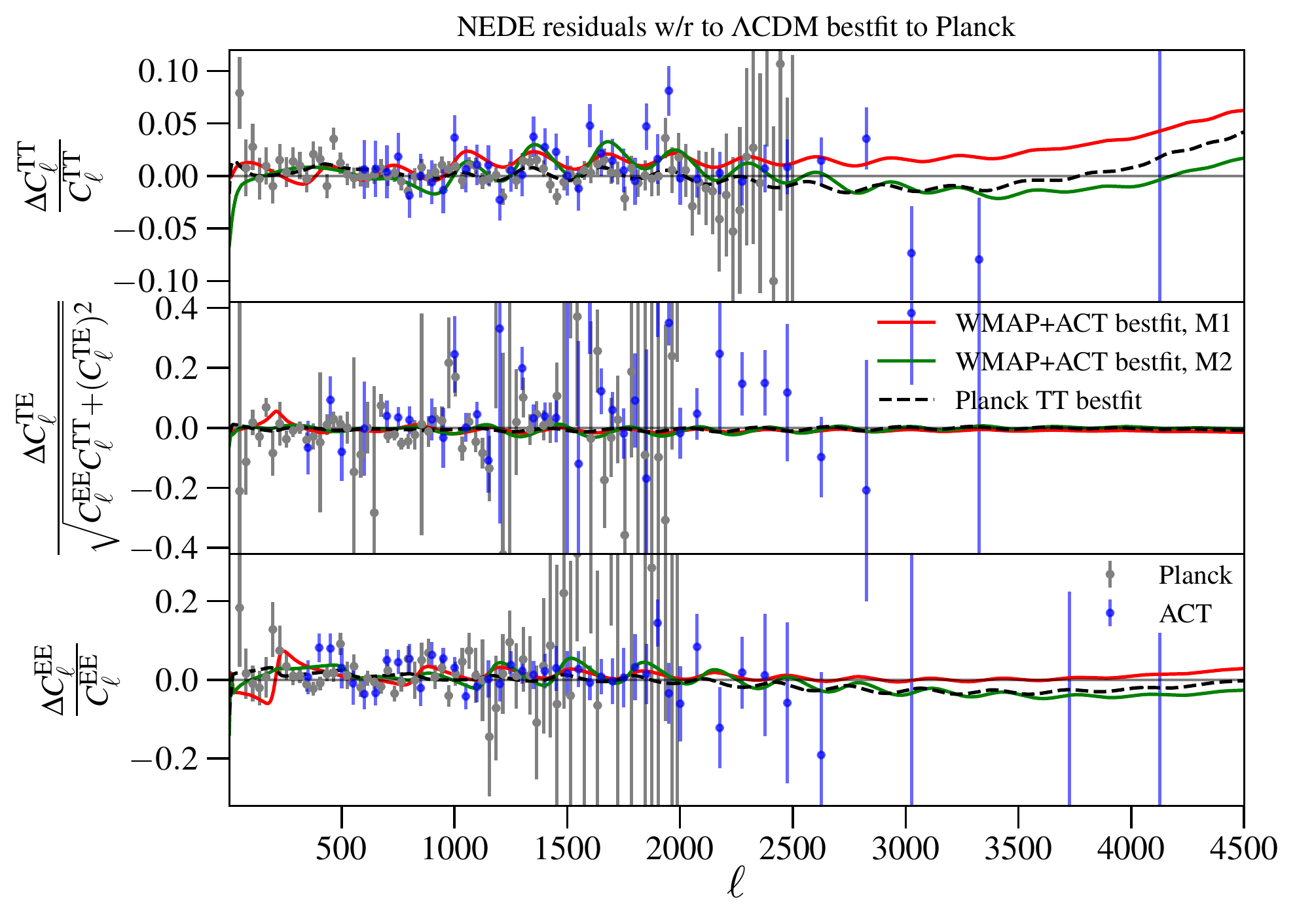}

    \caption{Same as Fig.~\ref{fig:axEDE_Cls}, for the NEDE case.}    \label{fig:NEDE_CLs}
\end{figure*}

\section{What is it about {\it Planck} data that disfavors EDE?}
\label{sec:Planck-vs-ACT}

\begin{figure}[!h]
    \centering
    \includegraphics[width=\columnwidth]{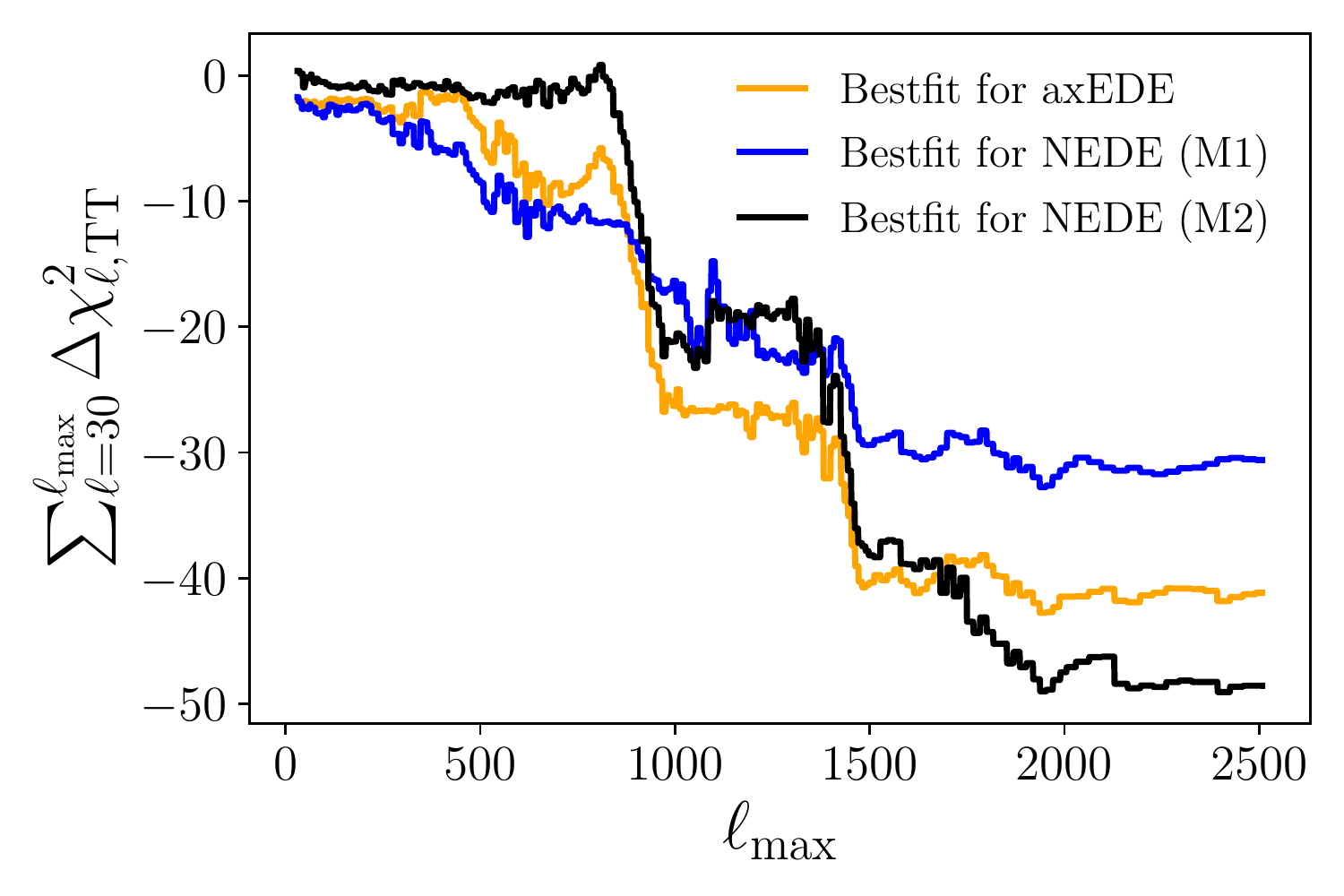}
    \caption{The relative, cumulative, $\Delta \chi^2_{\ell,\rm TT} \equiv \chi^2_{\rm EDE}\big|_{\rm {\it Planck}\ TT\ +\ Lowl}- \chi^2_{\rm EDE}\big|_{{\rm WMAP+ACT+}\tau}$. The drop around $\ell \sim 1000 - 1500$ indicates that the discrepancy between ACT and {\it Planck} as fit to both EDE models we have analyzed, occurs around this scale.}
    \label{fig:Dchi2}
\end{figure}

\begin{figure*}
\includegraphics[width=\columnwidth]{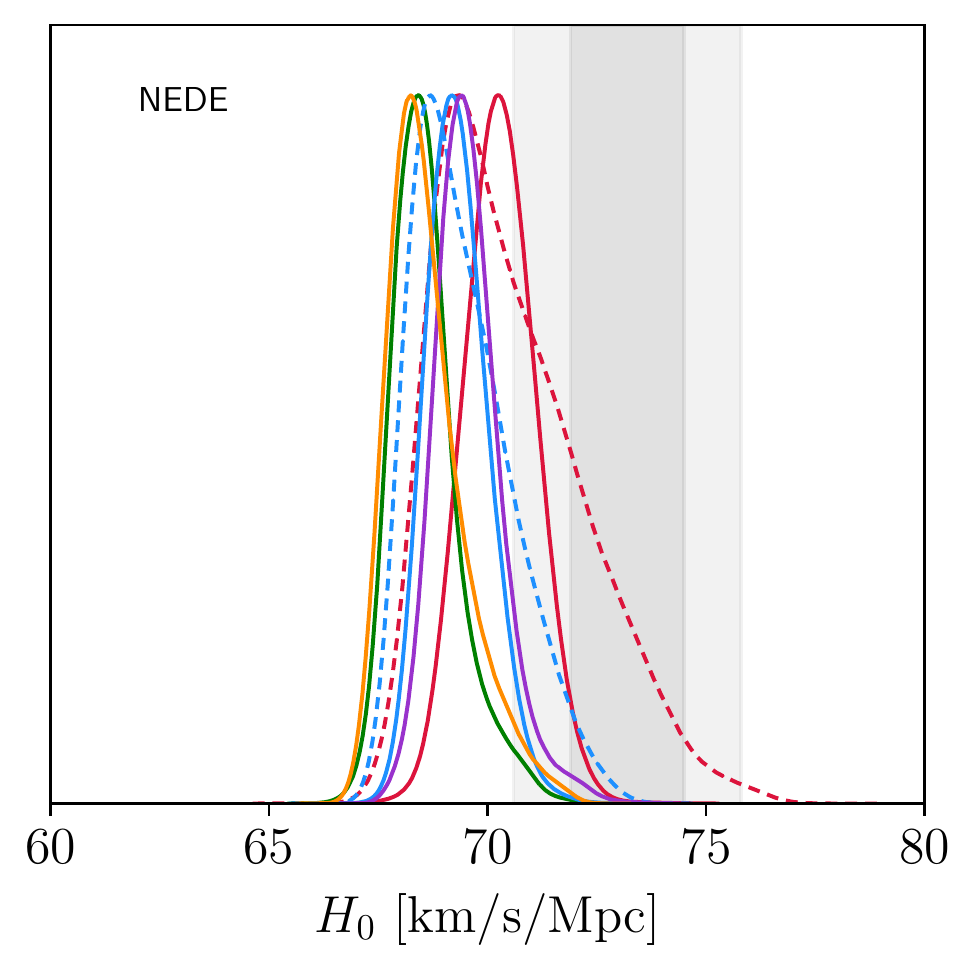}
\includegraphics[width=\columnwidth]{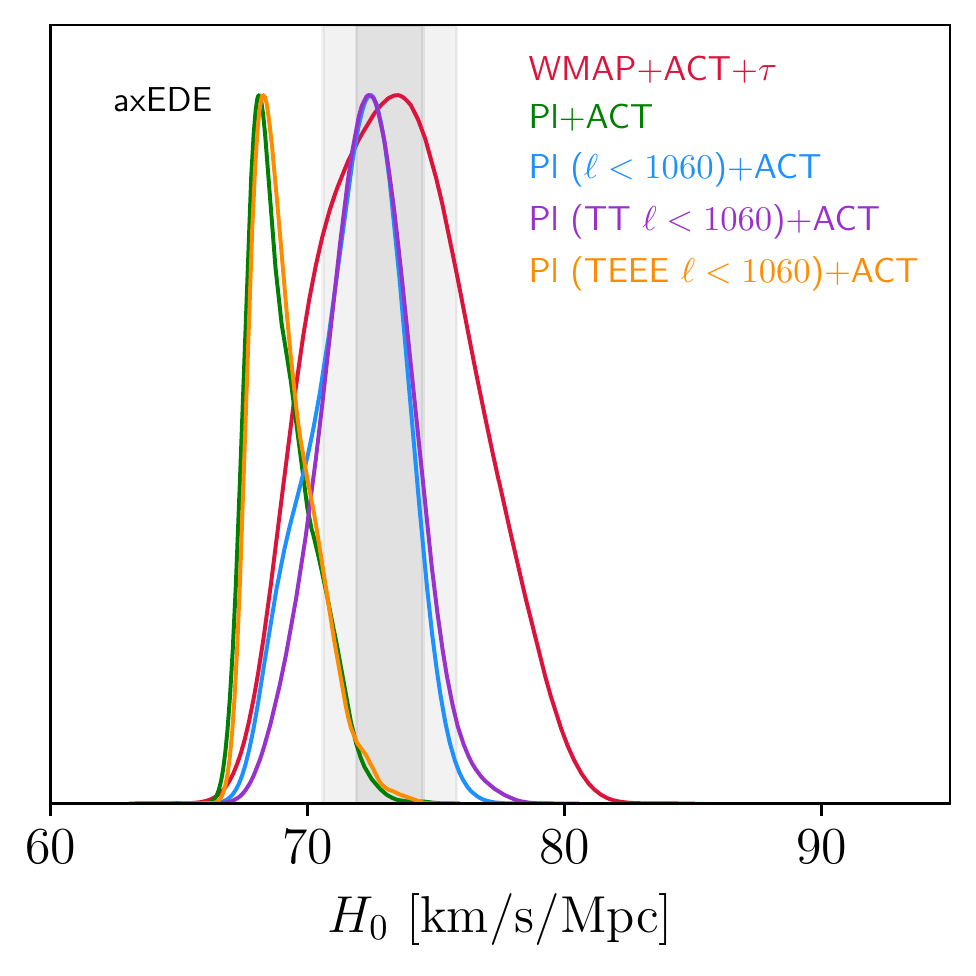}
   \caption{{\em Left panel:} 1D posterior distribution of $H_0$ in the axEDE (right panel) and NEDE (left panel) model when fit to {\it Planck} data restricted to $\ell < 1060$ together with ACT, compared to the results of the WMAP+ACT and full {\it Planck}+ACT analyses. In the NEDE case, we show the low-mass mode with a dashed line and the high-mass mode with a solid line.}
    \label{fig:axEDE_planckRestricted}
\end{figure*}
\subsection{$\Delta \chi^2$ considerations: the role of TT data}

To start to gauge what may be causing the difference in {\it Planck} vs.~ACT fits, Fig.~\ref{fig:axEDE_Cls} shows the residual of the CMB power spectra between the $\Lambda$CDM and EDE best-fit cosmologies of the WMAP+ACT+BAO+Pantheon data combination, compared to that of the  EDE best-fit cosmologies to {\it Planck} TT data alone. We also show the residuals of {\it Planck} and ACT data. In this case, we take the best-fit $\Lambda$CDM model to {\it Planck} high-$\ell$ TT,TE,EE+low-$\ell$ EE \cite{Aghanim:2018eyx} as a common reference model. One can see that the axEDE best-fit model from WMAP+ACT predicts a higher tail at high-$\ell$ than the {\it Planck} TT best-fit model, which (by eye) is consistent with the ACT data points but overshoots the {\it Planck} data at $\ell\gtrsim 1500$. The NEDE residuals tell a similar story, as shown in Fig.~\ref{fig:NEDE_CLs}, although the oscillatory behavior of the residuals in TT and EE makes it harder to isolate the feature responsible for constraining NEDE in {\it Planck}.

We can obtain a more precise statement of where the mismatch lies in the TT {\it Planck} power spectrum by exploring how the cumulative $\chi^2$ varies as a function of the maximum multipole. We show $\sum_{\ell = 30}^{\ell_{\rm max}}\Delta \chi^2_{\ell,\rm TT}$, 
where $\Delta \chi^2_{\ell,\rm TT}  \equiv \chi^2_{\rm min}(\rm {\it Planck}\ TT\ +\ lowl)- \chi^2_{\rm min}({{\rm WMAP+ACT+}\tau}$ )
as a function of $\ell_{\rm max}$ in Fig.~\ref{fig:Dchi2}. For the data sets WMAP+ACT+$\tau$, we fixed the cosmological parameters to the best-fit of these data and optimized the values of the {\it Planck} TT nuisance parameters. There we can see that that the $\sum_{\ell = 30}^{\ell_{\rm max}}\Delta \chi^2$ drops around $\ell \sim 1000$ and again around $\ell \sim 1500$ for both EDE models we have analyzed, confirming the qualitative intuition we gained from the residual plots in Fig.~\ref{fig:axEDE_Cls}.

\subsection{Analyses with restricted {\it Planck} data}

We can get additional insight on the constraining power of the features we isolated by `throwing away' parts of {\it Planck} data at high-$\ell$. We performed an MCMC analysis of {\it Planck} data together with ACT, now restricting the {\it Planck} $\ell$-range\footnote{We note that there is some overlap in $\ell$ between our restricted {\it Planck} data mimicking WMAP and ACT, which can potentially lead to double-counting of information. Yet, we note that the ACT team recommends no cut in $\ell$ when combining ACT with WMAP, and we therefore follow this recommendation when using restricted {\it Planck} data. Given that our main conclusions do not rely on restricted {\it Planck} data, we leave a more careful investigation to future work.} to $\ell < 1060$ (which also roughly matches the range of $\ell$ covered by WMAP\footnote{Though, note that WMAP noise increases significantly above $\ell \sim 500$, which makes the restricted {\it Planck} data more constraining than WMAP.} \cite{Huang:2018xle}). We compare removing information only in TT, only in TE and EE or in all data. In the left-panel (right-panel) of Fig.~\ref{fig:axEDE_planckRestricted}, we show the resulting 1D marginalized posterior distribution for $H_0$ for axEDE (NEDE). More detailed figures are presented in App.~\ref{app:resPlanck}.

Within the axEDE model, one can see that once the {\it Planck} TTTEEE data are restricted to the WMAP range, the combination of {\it Planck}+ACT leads to a shift in $H_0$ which is in good agreement with the WMAP+ACT results, with $H_0 = 71.93_{-1.5}^{+2.2}$ km/s/Mpc and $f_{\rm axEDE}(z_c) = 0.119_{-0.035}^{+0.065}$. We further isolate where this shift is coming from by focusing attention on the light-blue curve which just restricts the $\ell$-range of the {\it Planck} TT power spectrum. In this case the posterior distribution is nearly identical to the fully restricted {\it Planck} fit, indicating that TEEE data are not playing a role in constraining the axEDE model favored by ACT. On the other hand, when the full {\it Planck} TT  are included, but the TEEE data are restricted (orange curves), the constraints are compatible with the full {\it Planck} ones, providing evidence that there is a (slight) tension between the high-$\ell$ {\it Planck} and ACT power spectra.

Within the NEDE model, the bimodality in $m_{\rm NEDE}$ is still present when restricting {\it Planck} TTTEEE, albeit less well defined. Moreover, $H_0$ only reaches values around $69-70$ km/s/Mpc for both modes, while previously the M2 mode was compatible with $\sim 73$ km/s/Mpc. When including either the high $\ell$ {\it Planck} TT or TEEE data, the bimodality now disappears. The shift in $H_0$ in NEDE appears to depend on both the  high-$\ell$ temperature and polarization {\it Planck} power spectra.  When TT data are restricted, only the high-mass mode (with $H_0 \sim 70$) survives. When TEEE data are restricted, the TT data removes both modes, and the constraint is compatible with that of full {\it Planck}. This is another illustration of the fact that CMB data are highly sensitive to the dynamics of the EDE dynamics close to recombination.  We leave a more detailed analysis of a comparison between ACT and {\it Planck} constraints on NEDE to future work. 

We conclude that the strong constraints on the fraction of EDE arises mainly from {\it Planck} high-$\ell$ TT data, and that there are features in the {\it Planck} TT power spectrum which disfavor the WMAP+ACT axEDE cosmology around $\ell \sim 1000$ and $\ell \sim 1500$.  Similar features restrict the NEDE cosmology, while the TEEE data also impacts the ability of NEDE to reach high-$H_0$.

\section{Testing axEDE with mock data: is EDE artificially favored by WMAP+ACT?}
\label{sec:mock}

In this section, given similarities between the axEDE and NEDE model, we focus on the axEDE model, noting that our overall conclusions apply to both.

Having established a difference between the ACT and {\it Planck} TT power spectra when fitting to the axEDE model, we perform a mock data analysis to determine whether the preference for axEDE in the ACT data could be driven by some artificial complicated degeneracy within the multidimensional parameter space that would appear simply because ACT data are less precise than {\it Planck} in this multipole range. To that end, we perform analyses on two different mock data sets. 
First, we take the $\Lambda$CDM best-fit cosmology to {\em real} ACT data as the fiducial model, and fit both $\Lambda$CDM and axEDE to these mock data. Our goal is to check whether an artificial axEDE signal appears {\em even though $\Lambda$CDM is the `true' model}.
Second, we take the axEDE best-fit cosmology to real ACT data as the fiducial model, and perform the same set of analyses. In that case, our goal is to check to what extent the $\Lambda$CDM model can accommodate the axEDE signal by re-adjusting its parameters (and therefore lead to biased constraints relative to the fiducial model parameters). 

\begin{figure*}
    \centering
    \includegraphics[width=\columnwidth]{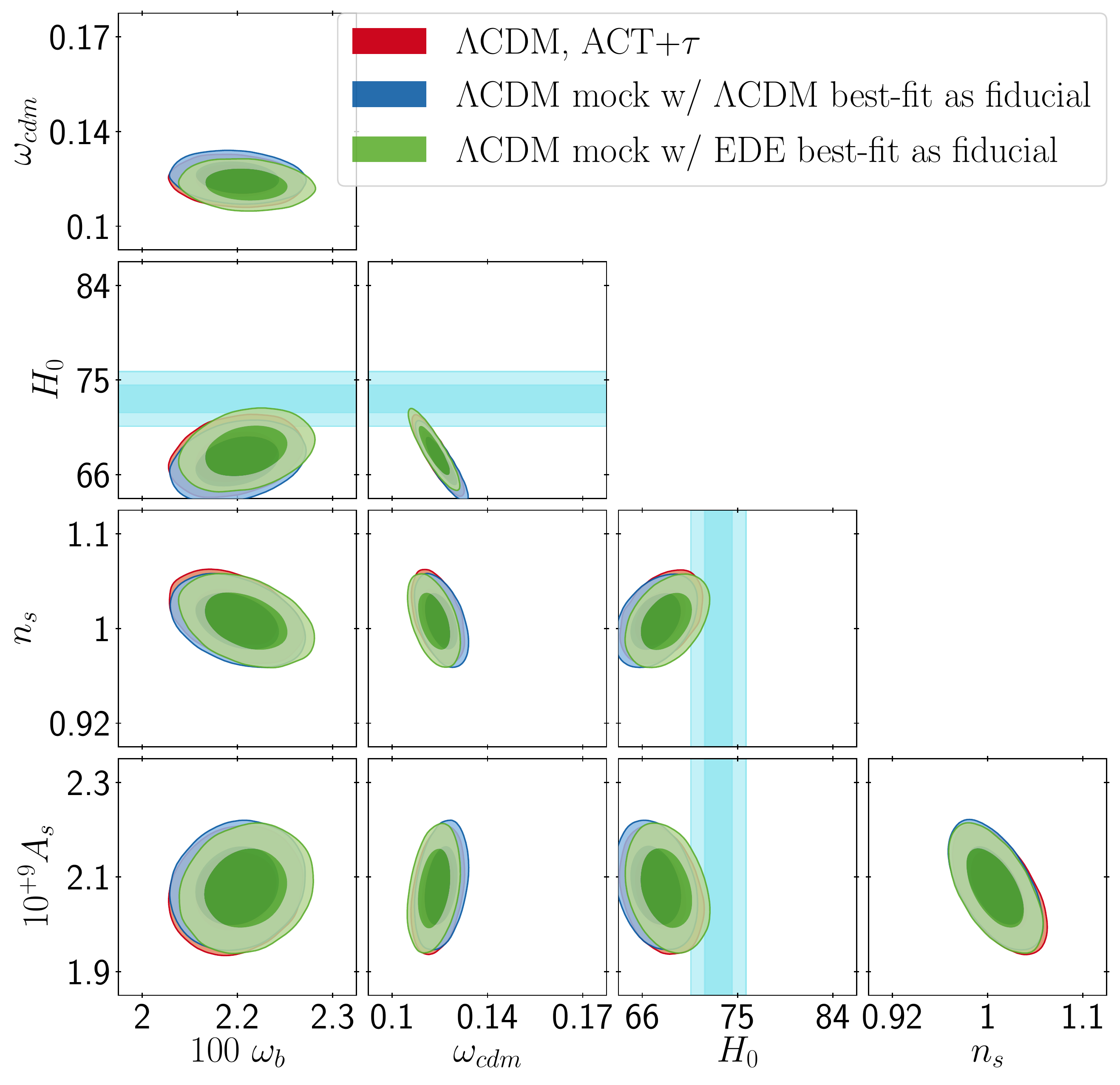}
    \includegraphics[width=\columnwidth]{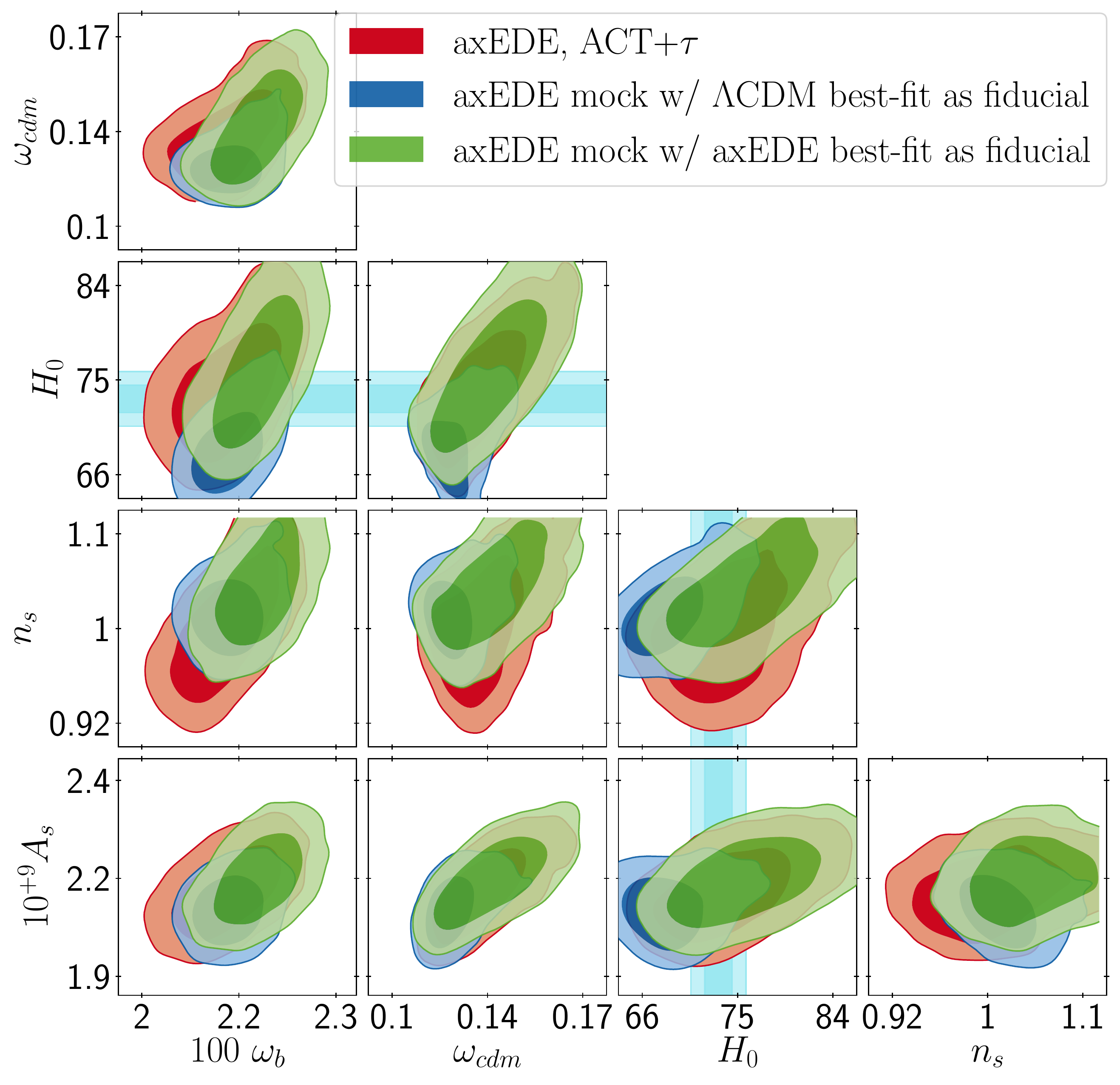}\\
        \includegraphics[width=\columnwidth]{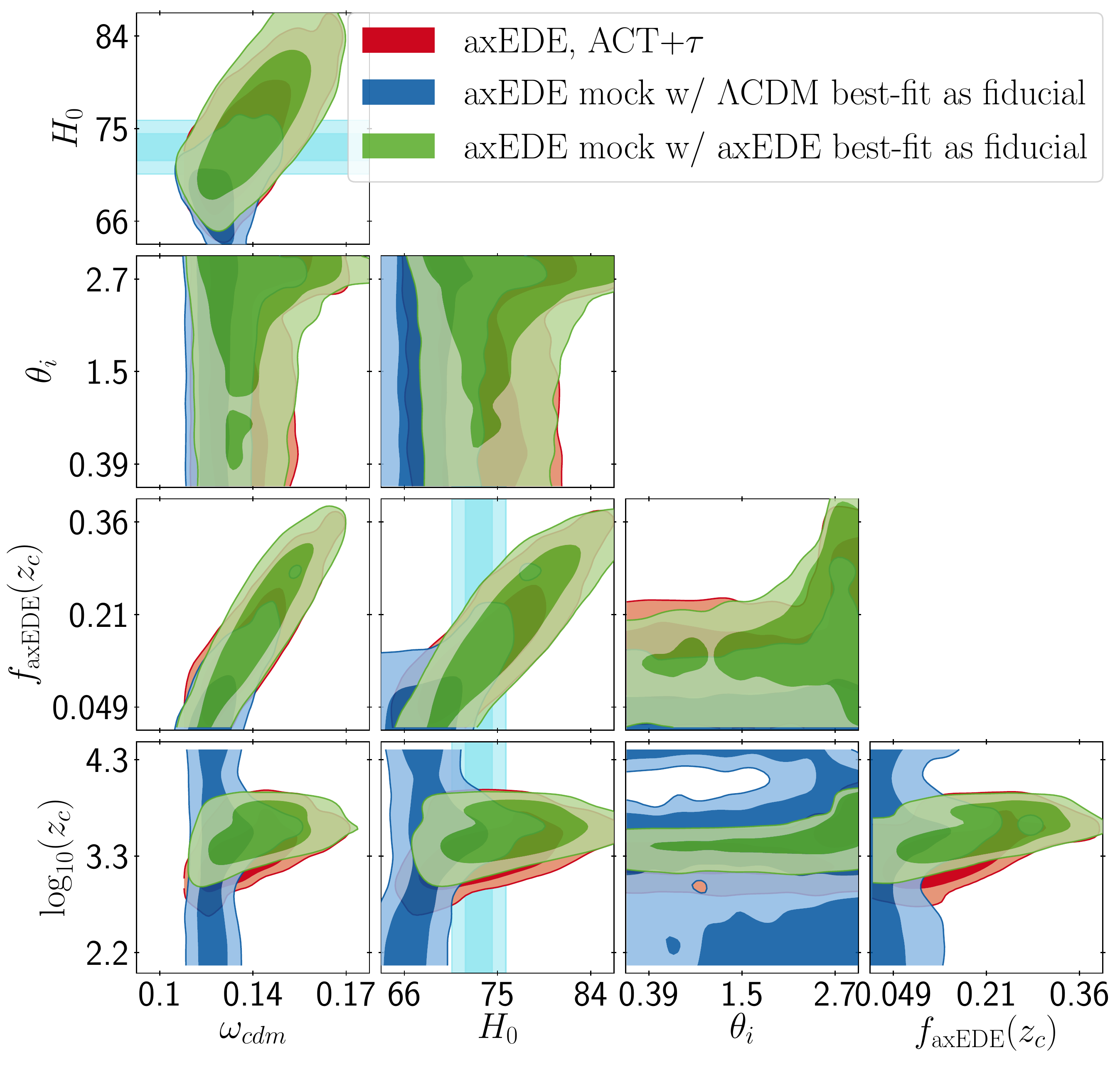}
    \caption{2D posteriors of a subset of parameters in the $\Lambda$CDM (top left) and axEDE (top right and bottom panel) cosmologies  fit to ACT data with a prior on $\tau$, compared with analysis of mock ACT data. In all panels, results of the real data analysis are shown in red. Results of the mock analysis with the $\Lambda$CDM best-fit model as fiducial are displayed in blue, while the results of the mock analysis with the axEDE best-fit model as fiducial are shown in green.}
    \label{fig:axEDE_mock}
\end{figure*}

In order to perform this mock analysis we compiled a modified likelihood code (based on the ACTPol Lite python likelihood) which uses the full ACTPol covariance matrix but allows for a different set of fiducial power spectra.  Fig.~\ref{fig:axEDE_mock} and Tables \ref{tab:real_ACT}, \ref{tab:fake_LCDM} and \ref{tab:fake_axEDE} show our results.

First, the axEDE cosmology provides an excellent fit to the \LCDM\ fiducial, with a $\Delta \chi^2 = 0.26$.  However, it does so simply because $\Lambda$CDM is nested within the axEDE cosmology, and no spurious axEDE signal is detected. The axEDE parameters $\log_{10}(z_c)$ and $\theta_i$ are unconstrained, while $f_{\rm axEDE}(z_c) < 0.162$. This is markedly different from their values when fit to the real data: $\log_{10}(z_c) = 3.31^{+2}_{-0.27}$, $f_{\rm axEDE}(z_c) = 0.152^{+0.054}_{-0.091}$. This  demonstrates that the apparent axEDE detection within ACT data cannot be attributed to a bias created by a lack of information compared to {\it Planck}.

Second, the \LCDM\ cosmology provides a decent fit to the axEDE fiducial, with $\Delta \chi^2 = 10.32$. It is interesting to note that this number is remarkably similar to what is obtained when comparing the \LCDM\ vs.~axEDE fit to the real ACT data, namely $\Delta \chi^2 = 9.3$. However, the posterior distributions for several parameters are significantly different from their fiducial values, showing biases up to $\sim 7\sigma$. For example, the axEDE fiducial values $H_0 = 77.49$ and $\omega_{\rm cdm} = 0.1460$ become $H_0 = 68.29 \pm 1.6$ and $\omega_{\rm cdm} = 0.1170 \pm 0.0038$. In fact, the marginalized constraints to the \LCDM\ parameters when fit to the axEDE fiducial are remarkably close to the values we find when fitting \LCDM\ to the actual ACT data. 

These results give us confidence that the axEDE fit to ACT data is not driven by a degeneracy in the data but is, in fact, consistent with the possibility that the data is better described by an axEDE cosmology than it is by \LCDM. 

\begin{table*}
 \scalebox{0.95}{
 \begin{tabular}{|l|c|c|} 
 \hline
  Data &  \multicolumn{2}{c|}{ACT+$\tau$ (real data)}  \\
          \hline
Model & $\Lambda$CDM & axEDE \\
 
        \hline
    \hline
  $f_{\rm axEDE}(z_c)$ & $-$ & $0.152(0.254)_{-0.091}^{+0.054}$\\
    $\log_{10}(z_c)$&$-$ & $3.31(3.78)_{-0.27}^{+0.2}$\\
    $\theta_i$ &  $-$ & unconstrained (2.99) \\
    \hline
 
        $H_0$ [km/s/Mpc]&$67.74(67.24)\pm1.6$  & $74.19(77.49)_{-4.8}^{+3.6}$ \\
    $100~\omega_b$ & $2.151(2.153)\pm0.031$& $2.161(2.244)_{-0.061}^{+0.055}$ \\
    $\omega_{\rm cdm}$&$0.1183(0.1196)\pm0.0039$ &$0.1349(0.1460)_{-0.015}^{+0.0081}$\\
    $10^{9}A_s$ &$2.072(2.084)\pm0.042 $ & $2.129(2.179)_{-0.07}^{+0.057}$\\
    $n_s$&$1.014(1.006)\pm0.016$ &  $1.002(1.064)_{-0.048}^{+0.032}$\\
    $\tau_{\rm reio}$&$0.0543(0.0511)\pm0.0074$ & $0.0546(0.0569)_{-0.0073}^{+0.0072}$\\
    \hline
    $\chi^2_{\rm min}$ &280.2 & 270.9\\
    \hline
        $\Delta \chi^2_{\rm min}$ &\multicolumn{2}{c|}{-9.3} \\

     \hline

\end{tabular} }
\caption{The mean (best-fit) $\pm 1\sigma$ errors of the cosmological parameters reconstructed in the $\Lambda$CDM and axEDE models from the analysis of real ACT data with a prior on $\tau=0.0543\pm0.0073$.}
\label{tab:real_ACT}
\end{table*}

\begin{table*}
 \scalebox{0.95}{
 \begin{tabular}{|l|c|c|c|} 
 \hline
  Data &  \multicolumn{3}{c|}{ACT+$\tau$ (mock data with $\Lambda$CDM fiducial)}  \\
          \hline
Model & fiducial model & $\Lambda$CDM  & axEDE \\
 
        \hline
    \hline
  $f_{\rm axEDE}(z_c)$ & $-$& $-$ & $ <0.162$ \\
    $\log_{10}(z_c)$&$-$&$-$ & unconstrained\\
    $\theta_i$ &  $-$ &$-$ & unconstrained \\
    \hline
 
        $H_0$ [km/s/Mpc]& 67.24& $67.26\pm1.6$  &$68.32_{-2.9}^{+1.9}$ \\
    $100~\omega_b$ &  2.153& $2.153\pm0.032$& $2.165_{-0.04}^{+0.034}$ \\
    $\omega_{\rm cdm}$& 0.1196&  $0.1196\pm0.0039$ &$0.1235_{-0.0071}^{+0.0038}$ \\
    $10^{9}A_s$ &2.084 & $2.081\pm0.042$ & $2.09_{-0.048}^{+0.045}$   \\
    $n_s$& 1.006& $1.006\pm0.016$  & $1.012_{-0.023}^{+0.018}$ \\
    $\tau_{\rm reio}$&0.0551  & $0.0545\pm0.0074$ & $0.0548_{-0.0074}^{+0.0072}$ \\
    \hline
    $\chi^2_{\rm min}$ & $-$& 0 & 0.2 \\

     \hline

\end{tabular} }
\caption{The mean (best-fit) $\pm 1\sigma$ errors of the cosmological parameters reconstructed in the $\Lambda$CDM and axEDE model from the analysis of mock ACT data generated from the $\Lambda$CDM best-fit cosmology extracted from the real ACT data, with a prior on $\tau=0.0543\pm0.0073$.}
\label{tab:fake_LCDM}
\end{table*}

\begin{table*}
 \scalebox{0.95}{
 \begin{tabular}{|l|c|c|c|} 
 \hline
  Data &  \multicolumn{3}{c|}{ACT+$\tau$ (mock data with axEDE fiducial)}  \\
          \hline
Model & fiducial model & $\Lambda$CDM &  axEDE \\
 
        \hline
    \hline
  $f_{\rm axEDE}(z_c)$ &  0.254& $-$ &$0.186_{-0.097}^{+0.099}$ \\
    $\log_{10}(z_c)$&3.78&$-$ & $3.52\pm0.23$\\
    $\theta_i$ &  2.9875&$-$ &$2.278_{-0.15}^{+0.82}$  \\
    \hline
 
        $H_0$ [km/s/Mpc]& 77.49& $68.29\pm1.6$ &  $76.18_{-5.3}^{+4.1}$    \\
    $100~\omega_b$ & 2.244 &$2.164\pm0.032$ &$2.215_{-0.046}^{+0.048}$   \\
    $\omega_{\rm cdm}$& 0.146&  $0.1170\pm0.0038$  & $0.1389_{-0.015}^{+0.012}$ \\
    $10^{9}A_s$ & 2.179& $2.076\pm0.043$  &$2.156_{-0.065}^{+0.061}$ \\
    $n_s$& 1.0642& $1.006\pm0.016$  &$1.032_{-0.035}^{+0.032}$ \\
    $\tau_{\rm reio}$&0.0569 &  $0.0551\pm0.0074$ & $0.0547_{-0.0075}^{+0.0073}$\\
    \hline
    $\chi^2_{\rm min}$ &$-$& 10.3 & 0 \\

     \hline

\end{tabular} }
\caption{The mean (best-fit) $\pm 1\sigma$ errors of the cosmological parameters reconstructed in the $\Lambda$CDM and axEDE model from the analysis of mock ACT data generated from the axEDE best-fit cosmology extracted from the real ACT data, with a prior on $\tau=0.0543\pm0.0073$.}
\label{tab:fake_axEDE}
\end{table*}

\section{Discussion and conclusions}
\label{sec:concl}

\subsection{Summary of the main findings}

In this paper, we have fit two models of EDE -- the phenomenological axion-like axEDE model from Ref.~\cite{Smith:2019ihp} and the `new' EDE (NEDE) model from Ref.~\cite{Niedermann:2019olb} -- to the latest data from the ACT collaboration \cite{Aiola:2020azj} in combination with data either from the WMAP or {\it Planck}, along with measurements of the BAO and uncalibrated SN1a from Pantheon data. 
Our work provides a clear example of how CMB measurements alone may discriminate between different EDE models. Our main results can be summarized as follows:

\begin{itemize}
    \item[\textbullet] ACT (with and without WMAP) prefers a non-zero EDE contribution at $\gtrsim 2 \sigma$, regardless of the model, {\em without the need to include any prior on $H_0$} (or $M_b$), and there is no residual tension between WMAP+ACT+BAO+Pantheon and SH0ES. This is in contrast with results from {\it Planck}+BAO+Pantheon, which only leads to an upper limit on the fraction of the EDE energy density at the critical redshift, for a (small) residual tension of $\sim 1.5\sigma$.
    \item[\textbullet] Yet, when conservatively combining ACT with {\it Planck} within the axEDE model, we find a {\em weaker} upper limit than from {\it Planck}-only, $f_{\rm axEDE}(z_c)<0.110$ (as opposed to $f_{\rm axEDE}(z_c)<0.084$). Remarkably, there is no tension between {\it Planck}+ACT+BAO+Pantheon and SH0ES ($0.4\sigma$), in stark contrast with $\Lambda$CDM for which the tension is $4.5\sigma$. However, the NEDE model is more strongly constrained than without ACT, yielding $f_{\rm NEDE}(z_*)<0.082$ (as opposed to $f_{\rm NEDE}(z_*)<0.116$)  and a residual $2.9\sigma$ tension. This shows that the combination of {\it Planck} and ACT can break degeneracies between EDE models.
    \item[\textbullet] Within the axEDE model, it is interesting to note that the ACT and WMAP+ACT 
    do not place any constraint on the initial field displacement, $\theta_i$. However, when we replace WMAP with the {\it Planck} data restricted to $\ell<1060$ we find a relatively tight constraint, $\theta_i = 2.705_{-0.067}^{+0.2}$. This indicates that the more precise measurements from {\it Planck} at lower multipoles are sensitive to the details of the perturbative EDE dynamics \cite{Smith:2019ihp}. Additionally,  WMAP+ACT seem to favor a critical redshift $z_c$ for the axEDE transition slightly lower than matter-radiation equality redshift $z_{\rm eq}$, although a combined analysis of Planck+ACT+SH0ES does favor $z_c$ around $z_{\rm eq}$.
     \item[\textbullet]Similarly, in the NEDE model, the trigger-field mass (which controls the redshift of the transition) favored by `{\it Planck}+SH0ES' is different from (although compatible with) that from the WMAP+ACT analysis, which show a bimodal distribution. 
    \item[\textbullet] We have shown that restricting {\it Planck} data $\ell$-range to $\ell < 1060$ (which roughly matches the range of $\ell$ covered by WMAP), in combination with ACT, leads to similar results as the WMAP+ACT analysis. We have further identified that there are differences between ACT and Planck TT spectra around $\ell\sim 1000$ and $\ell \sim 1500$, where ACT data are systematically higher than that from {\it Planck}, which restrict the EDE contribution within {\it Planck} data.
    \item[\textbullet] From the analysis of mock ACT data, we have confirmed that the preference for a non-zero axEDE contribution does not come from a lack of low to intermediate angular scale information which would bias the marginalized posteriors. Assuming the $\Lambda$CDM best-fit cosmology as fiducial model, we have shown that the axEDE posteriors are compatible with $f_{\rm axEDE}=0$ at $1\sigma$, in stark contrast with the result of the real data analysis.  On the other hand, with an axEDE fiducial, we found that the reconstructed posteriors in the $\Lambda$CDM model show bias up to $\sim 7\sigma$ compared to their fiducial value (most notably for $\{H_0,\omega_{\rm cdm}\}$), at a cost in $\chi^2_{\rm min}=+10$. Remarkably, the reconstructed parameters and the $\chi^2_{\rm min}$ penalty is in very good agreement with the real data analysis from ACT. This further indicates that (contrary to {\it Planck}) ACT data seem to (mildly) favors the axEDE cosmology over the standard $\Lambda$CDM model (and by extension, other EDE models).
\end{itemize}

\subsection{From the Hubble tension to a CMB tension?}

\begin{figure}
    \centering
    \includegraphics[width=0.75\columnwidth]{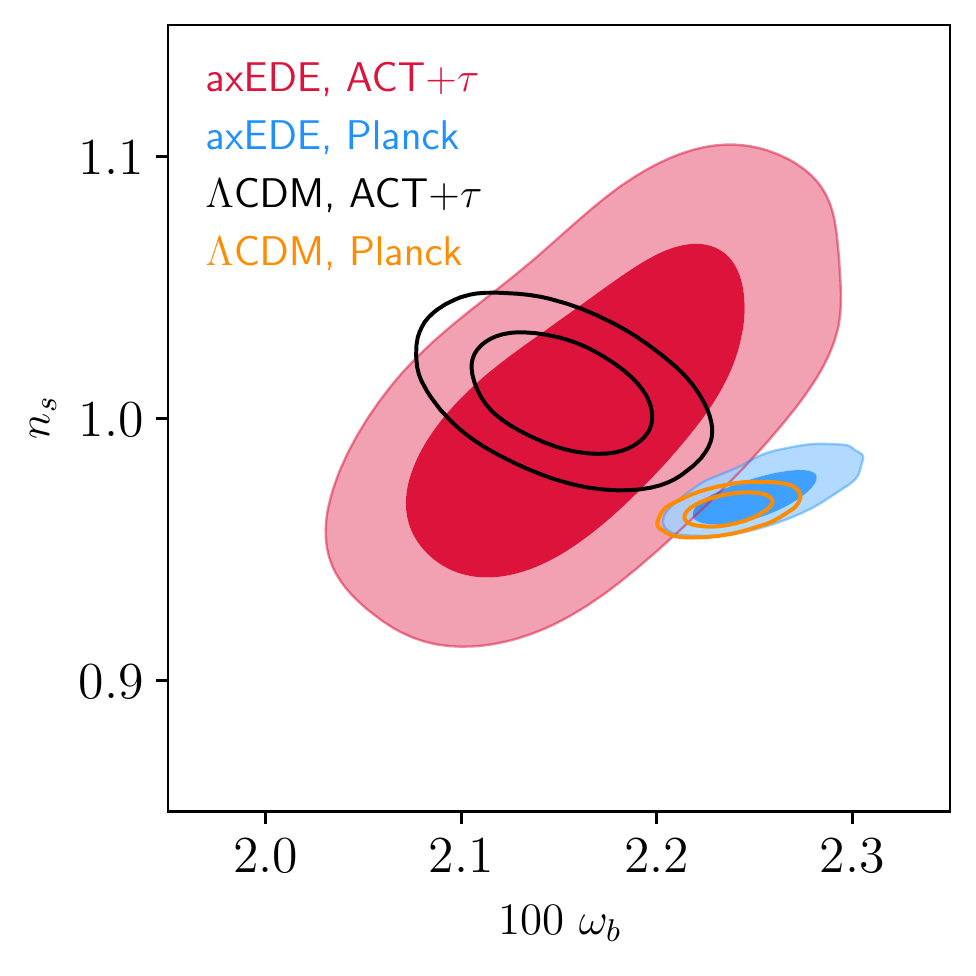}
    \caption{2D posteriors in the $n_s$ vs.~$\omega_b$ plane for various CMB observations fit to an axEDE cosmology. Unlike in \LCDM\, the degeneracy is positive for both low and high-$\ell$ observations.}
    \label{fig:omegab_vs_ns}
\end{figure}

\begin{figure*}
    \centering
    \includegraphics[width=2.1\columnwidth]{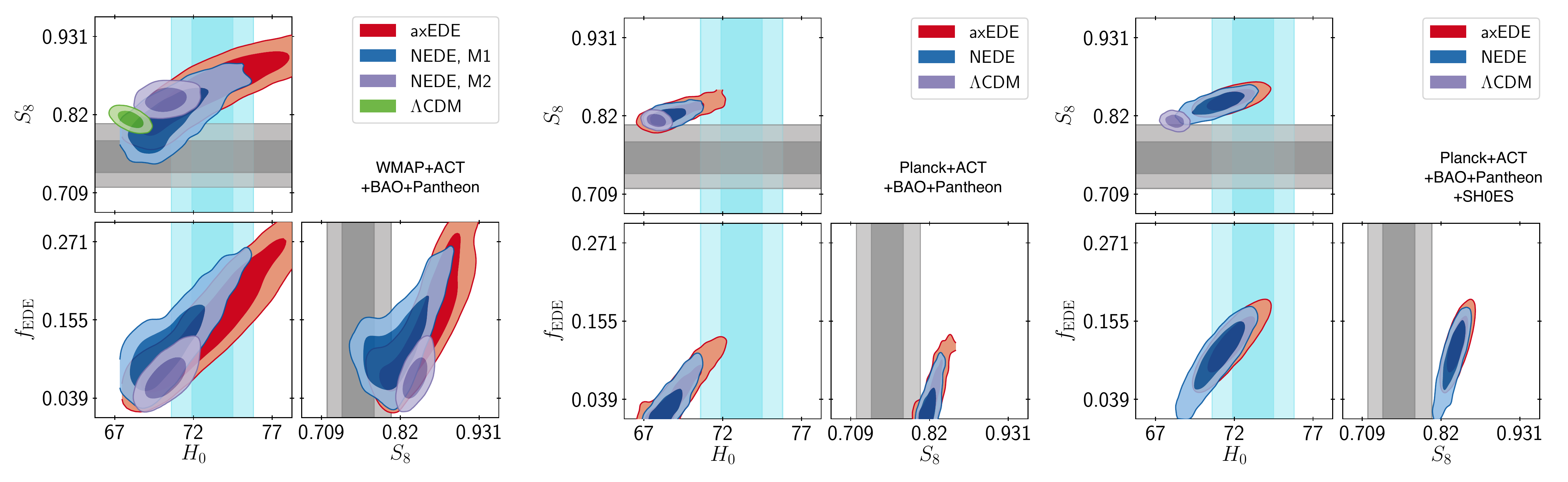}

 \caption{{\em Summary plots.} {\em Left panel:} 2D posteriors of $\{f_{\rm EDE}(z_c), H_0, S_8\}$ for the axEDE and NEDE models when fit to WMAP+ACT+BAO+Pantheon data, without any information from SH0ES. We also show the $\Lambda$CDM case for comparison. Note that the NEDE model features two different modes for low (M1) and high (M2) trigger field mass, both represented here.
 {\em Middle panel:} Same as left panel, in the Planck+ACT+BAO+Pantheon case.
 {\em Right panel:} Same as middle panel,  now including SH0ES.}
 \label{fig:axEDE_concl}.
\end{figure*}
Previous analyses have identified other potential tensions between the ACT, WMAP, and {\it Planck}. Within the context of $\Lambda$CDM, Ref.~\cite{Aiola:2020azj} used the posterior distributions for $n_s$ and $\omega_b$, to argue that the high-$\ell$ ACT data may be in slight tension with WMAP and {\it Planck}. They point out that a five percent decrease in the ACT TE calibration would shift the constraints into statistical agreement, but noted that there is no reason to introduce such a `correction'. This `tension' was mentioned in Ref.~\cite{Lin:2020jcb} when demonstrating that a fit of `acoustic dark energy' \cite{Lin:2019qug} to {\it Planck}+ACT is constrained by the intermediate scale ($\ell \sim 500$) {\it Planck} polarization.

It has also been noted that constraints on \LCDM\ using measurements of the CMB E-mode polarization prefer a value of $H_0$ that is higher than the value obtained from analyses that includes temperature data \cite{2016ApJS..227...21T,Aghanim:2018eyx,SPT:2017jdf,Aiola:2020azj}. Recently, Ref.~\cite{Addison:2021amj} systematically explored why the polarization and temperature results differ and concluded that it is mainly driven by different degeneracy directions in the $\{ \omega_b,n_s\}$ plane between high and low-$\ell$ CMB measurements. The joint constraints naturally break these degneracies, leading to the increase in the inferred value of $\omega_b$ and, as a result, a lower value of $H_0$.

We find that when fitting the axEDE model to ACT data the degeneracy between $\omega_b$ and $n_s$ is altered. In particular, the negative correlation between the amplitude of the small-scale power spectrum and $\omega_b$ (through Silk damping) is broken by the presence of an EDE phase. A comparison between constraints to these parameters using ACT vs.~{\it Planck} data is shown in Fig.~\ref{fig:omegab_vs_ns}. There we can see that the axEDE fit is more consistent than \LCDM, reducing the previously reported temperature vs.~polarization tension. 

The ACT/{\it Planck} tension that we identify is different than those that have been previously reported in the literature. We find that the $\ell \gtrsim 1000$ {\it Planck} TT power spectrum may be in tension with the ACT power spectra. This is reminiscent of the investigations of the consistency of cosmological parameters measured from high and low multipoles from Planck \cite{Addison:2015wyg,Planck:2013pxb,Planck:2019nip}. Although within the context of \LCDM\ these tensions are not statistically significant \cite{Planck:2019nip}, this work raises the possibility that when analyzed with other cosmological models high and low temperature multipoles may be in tension. We leave a more systematic exploration of this possibility to future work. 

\subsection{Final thoughts}

These `hints' for the presence of EDE should certainly be interpreted with care, given that they depend on the data-combination, and further work needs to be done to establish whether they are real. For instance, analysis with SPT-3G data would be interesting to test whether these results are `ground-based experiment (in)dependent', or whether they could come from fluctuation in a given patch of the sky. In addition to this, the ACT collaboration already has data beyond DR4 (i.e., DR5 \cite{Mallaby-Kay:2021tuk} which includes data acquired in 2017–18) which may also shed light on the DR4 preference for EDE cosmologies.

Moreover, it should be noted that, while the reconstructed cosmology in the EDE models leads to SN1a intrinsic magnitude $M_b$ (and Hubble rate today $H_0$) compatible with local measurements, the $S_8$ tension increases. This is illustrated in Fig.~\ref{fig:axEDE_concl}, where we compare the predicted $S_8$ in EDE cosmologies that resolve the $H_0$ tension with the measurement from KiDS+BOSS+2dFLenS \cite{Heymans:2020gsg}. Concretely, in the axEDE cosmology resulting from the Planck+ACT+BAO+Pantheon+SH0ES data combination, we find (Gaussian) tension at the $3.3\sigma$ level with the value from KiDS+BOSS+2dFLenS \cite{Heymans:2020gsg}, and $3.2\sigma$ with that from DES `3x2pt' statistics \cite{DES:2021wwk}. In comparison, against the same data combination, the $\Lambda$CDM leads to $2.3\sigma$ and $2.1\sigma$ tension with respect to KiDS+BOSS+2dFLenS and DES respectively\footnote{Dropping the SH0ES likelihood, in the $\Lambda$CDM model we find $2.6\sigma$ and $ 2.5\sigma$  tension with respect to KiDS+BOSS+2dFLenS and DES respectively.}. Similar issues were raised in the literature \cite{Hill:2020osr,Ivanov:2019pdj,DAmico:2020ods}, but it was shown that EDE models are currently not excluded by LSS data \cite{Murgia:2020ryi,Smith:2020rxx}. Future measurements of the halo mass function at high-$z$ will provide an important test of EDE cosmologies \cite{Klypin:2020tud}. Nevertheless, resolving the $S_8$ tension requires different physics than that at play in resolving the $H_0$ tension, namely one must decrease the matter power spectrum at scales $k\sim 0.1-1$ Mpc$/h$ \cite{Lange:2020mnl}. Numerous models have been proposed to that end (involving e.g. hot dark matter \cite{Das:2021pof}, fuzzy dark matter \cite{Lague:2021frh,Allali:2021azp}, CDM decays to warm daughters \cite{Abellan:2020pmw,Abellan:2021bpx}, or CDM interaction with a new `dark radiation' component \cite{Lesgourgues:2015wza,Buen-Abad:2017gxg,Buen-Abad:2018mas,Archidiacono:2019wdp,Heimersheim:2020aoc} or dark energy \cite{DiValentino:2019ffd,Lucca:2021dxo}) and, while perhaps theoretically unpleasing in light of `Occam's razor', could resolve the $S_8$-tension independently of EDE. As a concrete example, it was recently shown that the combination of a ultra-light axion with mass $m\sim 10^{-28}$ eV contributing at $5\%$ to the CDM density, together with the NEDE model studied here, could resolve both tensions \cite{Allali:2021azp}.
\begin{figure}
    \centering
    \includegraphics[width=\columnwidth]{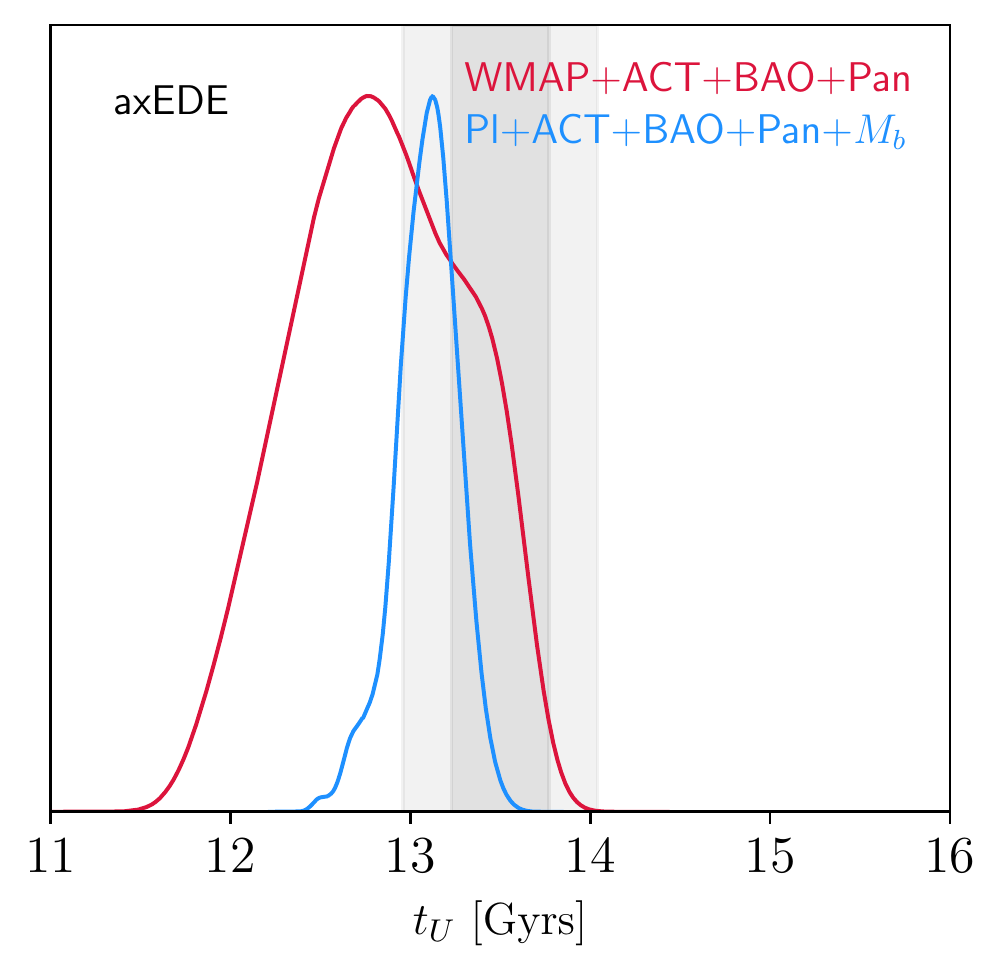}
    \caption{1D posterior distribution of the age of the universe $t_U$ reconstructed in the axEDE cosmology when fit to either WMAP+ACT+BAO+Pantheon or Planck+ACT+BAO+Pantheon+$M_b$. The grey band represents the globular cluster estimates of $t_U=13.5\pm0.27$ Gyrs. }
    \label{fig:EDE_age}
\end{figure}

Another (potential) issue raised by EDE cosmologies, and more generally models resolving the Hubble tension by adjusting the sound horizon, is that the Universe is notably younger than in $\Lambda$CDM, which could lead to tensions with the measured age of old objects such as globular clusters of stars (GC) \cite{Bernal:2021yli,Boylan-Kolchin:2021fvy,Vagnozzi:2021tjv}. Recently, such measurements have been shown to be in slight tension with the prediction from the axEDE cosmology resulting from a fit to {\it Planck}+SH0ES \cite{Bernal:2021yli,Boylan-Kolchin:2021fvy}. In our case, we illustrate this potential tension in Fig.~\ref{fig:EDE_age} for the axEDE model reconstructed from our WMAP+ACT and Planck+ACT+$M_b$ analyses (two cosmologies for which the Hubble tension is resolved), where we also display the recent measurements from GC, $t_{\rm U}=13.5\pm0.027$ Gyrs \cite{Valcin:2020vav,Valcin:2021jcg,Bernal:2021yli}. We find that, while the predicted age of the universe is systematically lower than that measured from GC, the differences are not statistically significant (between $0.9\sigma$ and $1.2\sigma$ agreement for the cosmologies resulting from analyses that include ACT data). Similar considerations apply to `cosmic chronometers' (e.g.~Ref.~\cite{Vagnozzi:2020dfn}) which are currently too imprecise to probe the EDE scenarios \cite{Schoneberg:2021qvd}, but could provide interesting tests in the future.

While this tentative result from ACT shows that there is a path to detecting an EDE in CMB data alone (as advocated in Ref.~\cite{Smith:2019ihp}), more accurate measurements of the CMB power TT and EE spectra above $\ell\sim 1000$, as well as around $\ell \sim 300-500$ in EE, with surveys such as the Simons Observatory \cite{SimonsObservatory:2018koc} and CMB-S4 \cite{CMB-S4:2016ple}, will play a crucial role in firmly establishing (or excluding) the presence of dark energy at early times in the Universe, and help in differentiating between models.

{\it Note added:} During the final stages of preparing this manuscript we became aware of Ref.~\cite{Hill:2021yec}, which also investigates how ACT data is fit by the phenomenological axion-like axEDE model. Our conclusions, as they relate to axEDE, are very similar to theirs. They go further in their analysis of the preference for EDE within ACT data, clearly establishing that it is driven by the low-$\ell$ EE multipoles. However, our work also explores the NEDE model, investigates a little further the source of tension between ACT and {\it Planck} fits, and performs a mock analysis to establish that the preference for axEDE in ACT data is not the result of a bias due to complicated degeneracy appearing when {\it Planck} data are removed from the analysis.

\acknowledgments

We thank Graeme Addison and Julien Lesgourgues for their help regarding the ACT likelihood, the mock data analysis and comments about our Planck-restricted analysis. We thank Florian Niedermann and Martin Sloth for helpful comments regarding the NEDE model on an earlier version of this draft. We also thank Eleonora Di Valentino, Tanvi Karwal, Marc Kamionkowski, Antony Lewis, Meng-Xiang Lin, Joel Primack, and Adam Riess, for helpful conversations.  TLS is supported by NSF Grant No.~2009377, NASA Grant No.~80NSSC18K0728, and the Research Corporation. AB is supported by NSF Grant No.~2009377 and the Provost's office at Swarthmore College. This work used the Strelka Computing Cluster, which is run by Swarthmore College. We acknowledge the use of the Legacy Archive for Microwave Background Data Analysis (LAMBDA), part of the High Energy Astrophysics Science Archive Center (HEASARC). HEASARC/LAMBDA is a service of the Astrophysics Science Division at the NASA Goddard Space Flight Center.  This work has been partly supported by the CNRS-IN2P3 grant Dark21. The authors acknowledge the use of computational resources from the Dark Energy computing Center funded by the Excellence Initiative of Aix-Marseille University - A*MIDEX, a French "Investissements d'Avenir" programme (AMX-19-IET-008 - IPhU).

\bibliography{bibliography}

\begin{appendix}

\section{Full ACTPol likelihood}
\label{ap:full}

\begin{figure}[h!]
    \centering
    \includegraphics[width=\columnwidth]{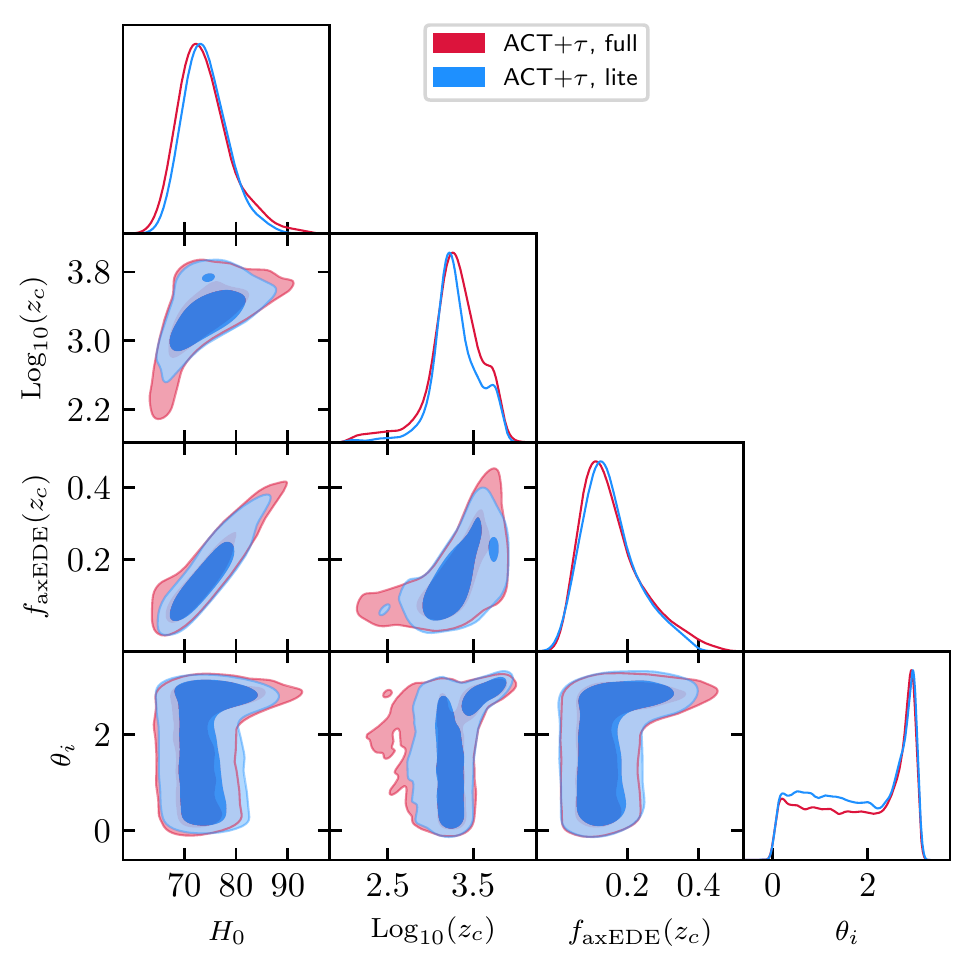}
    \caption{A comparison between the 2D posteriors of $\{H_0,f_{\rm axEDE}(z_c),\log_{10}(z_c),\theta_i\}$ reconstructed with the full ACT likelihood (red) or the `lite version' (blue). }
    \label{fig:full-vs-lite}
\end{figure}

Throughout this paper we have used the ACTPol Lite likelihood \cite{Aiola:2020azj} which marginalizes over a number of frequency-dependent nuisance parameters. In order to be sure that the axEDE parameters are uncorrelated with these additional nuisance parameters we implemented a modified version of CAMB \cite{Lewis:1999bs}, which incorporates the axEDE dynamics, and analyzed this model using the full, frequency dependent, ACTPol likelihood \cite{ACT:2020frw} using \texttt{CosmoMC} \cite{Lewis:2002ah,Lewis:2013hha}. We used the same priors (as listed in Ref.~\cite{ACT:2020frw}) as the standard ACT analysis. As shown in Fig.~\ref{fig:full-vs-lite}, the resulting marginalized constraints on the axEDE parameters are not changed when using the full ACT likelihood, compared to the constraints with the `lite' version, justifying our use of the ACTPol Lite likelihood when exploring constraints to axEDE. Given the similarity between axEDE and NEDE extensions of \LCDM, it is reasonable to expect that the NEDE constraints will also be unchanged when using the full likelihood. 

\section{Triangle plots comparing WMAP and restricted {\it Planck}}
\label{app:resPlanck}

\begin{figure*}[ht]
    \centering
    \includegraphics[width=\columnwidth]{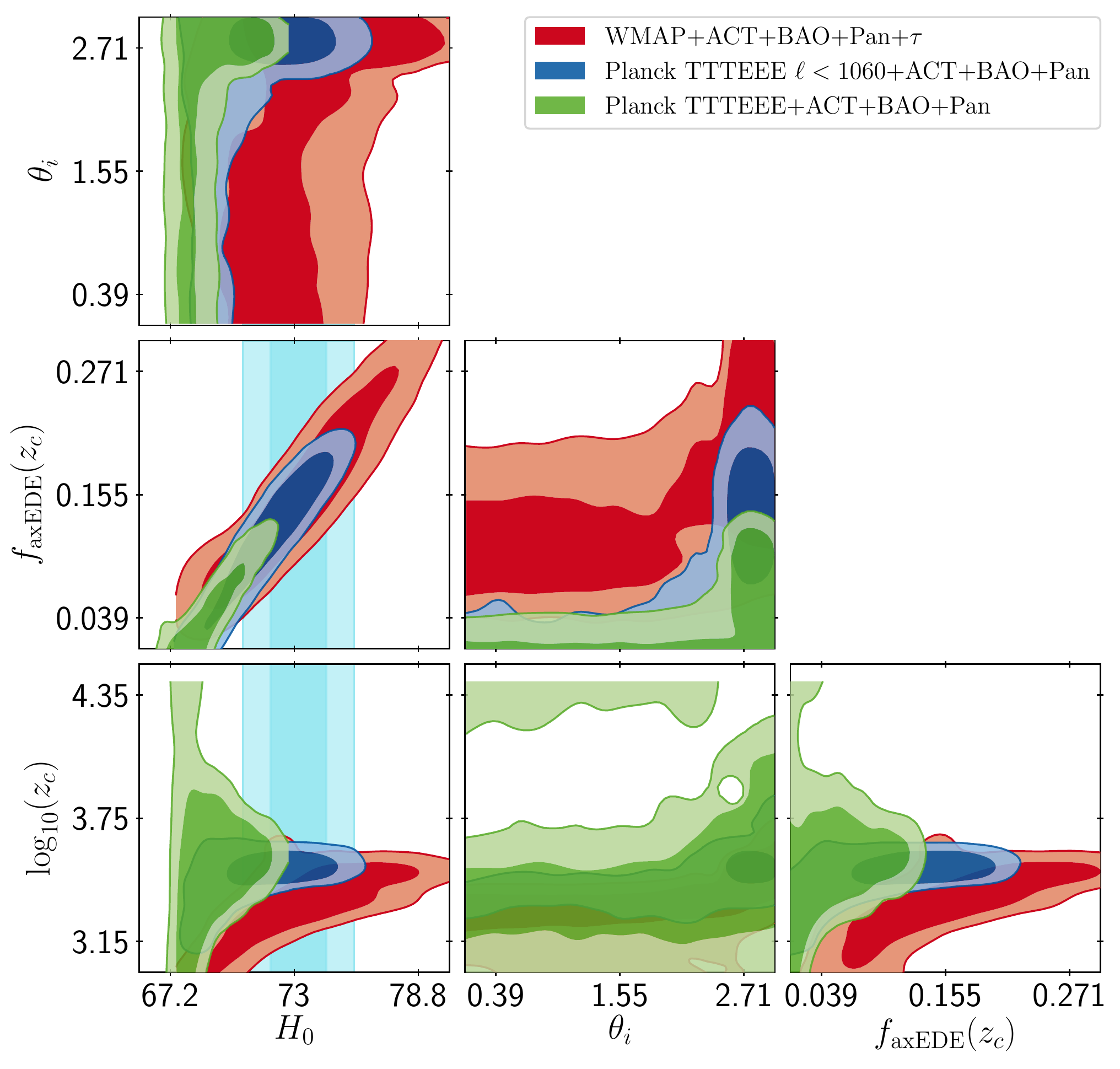}
     \includegraphics[width=\columnwidth]{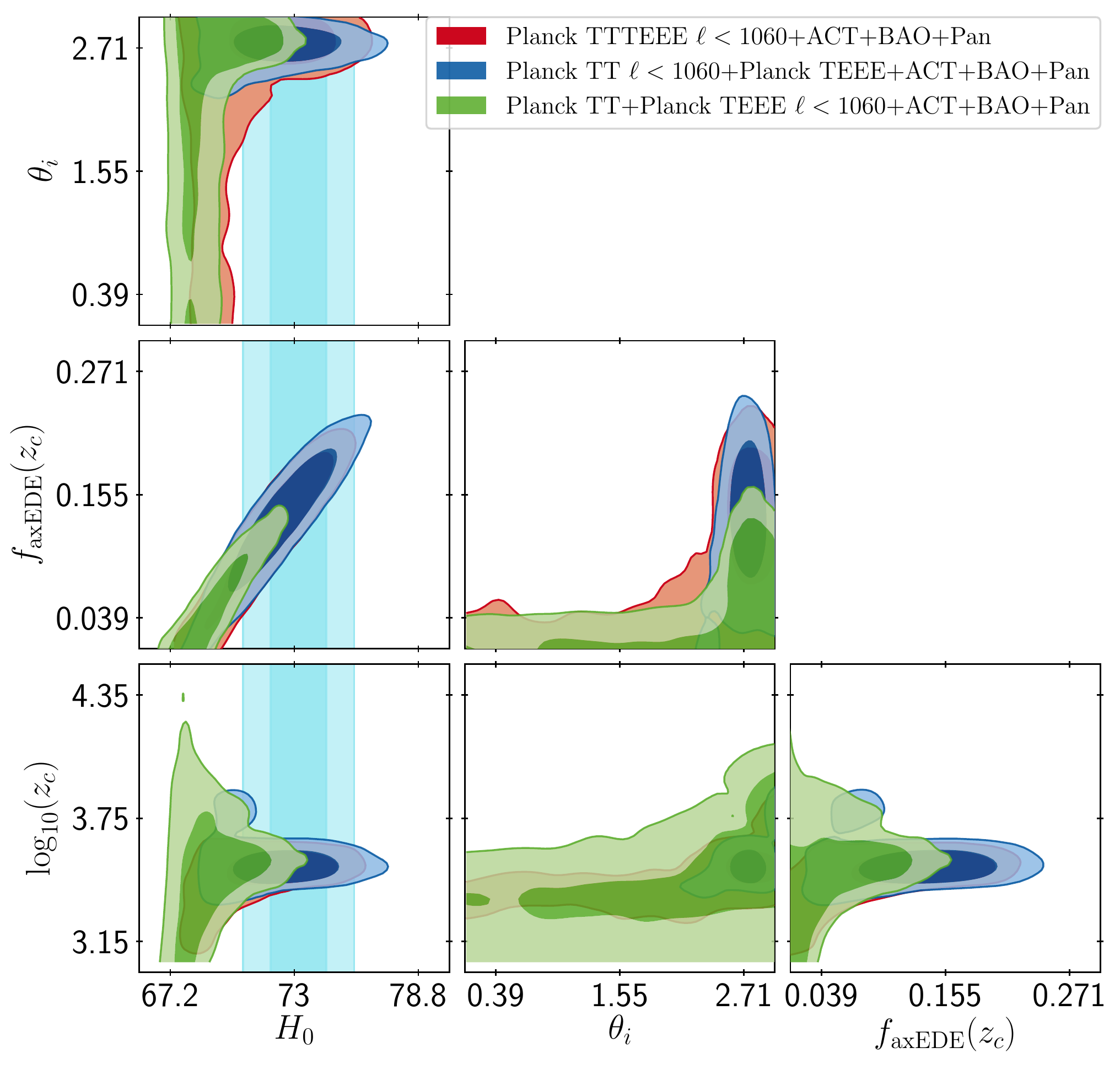}

    \caption{2D posteriors of a subset of parameters in the axEDE cosmology fit to various combinations of data. In the left panel, we compare the posteriors obtained when analyzing ACT together with either WMAP, {\it Planck} data restricted to $\ell<1060$, or the full multipole range of {\it Planck}. In the right panel, we show the difference between restricting {\it Planck} in TT, TEEE or both. All analyses also include BAO and Pantheon data.}
    \label{fig:axEDE_full_PlanckRes}
\end{figure*}

\begin{figure*}[ht]
    \centering   
    \includegraphics[width=1\columnwidth]{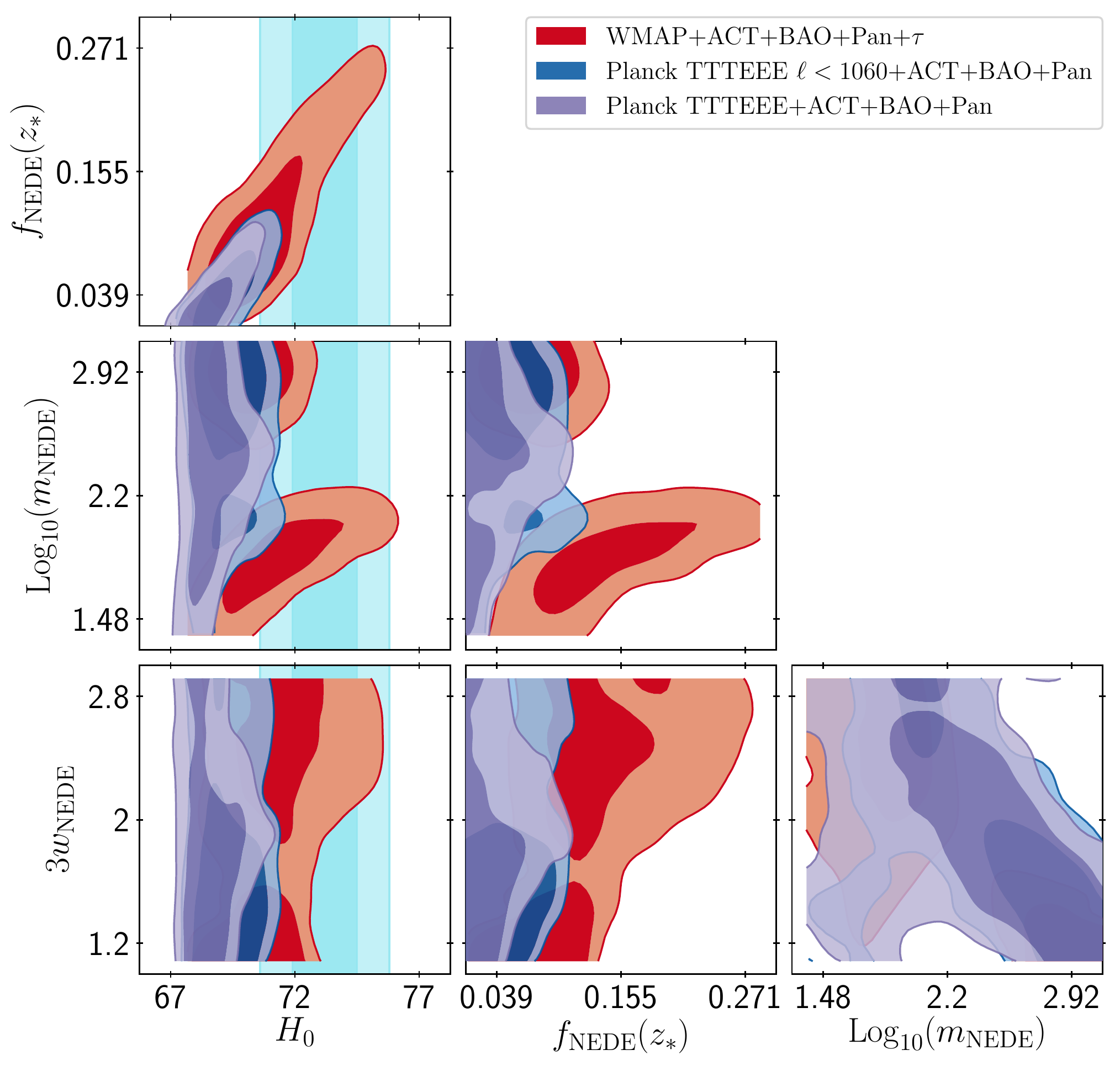}
    \includegraphics[width=1\columnwidth]{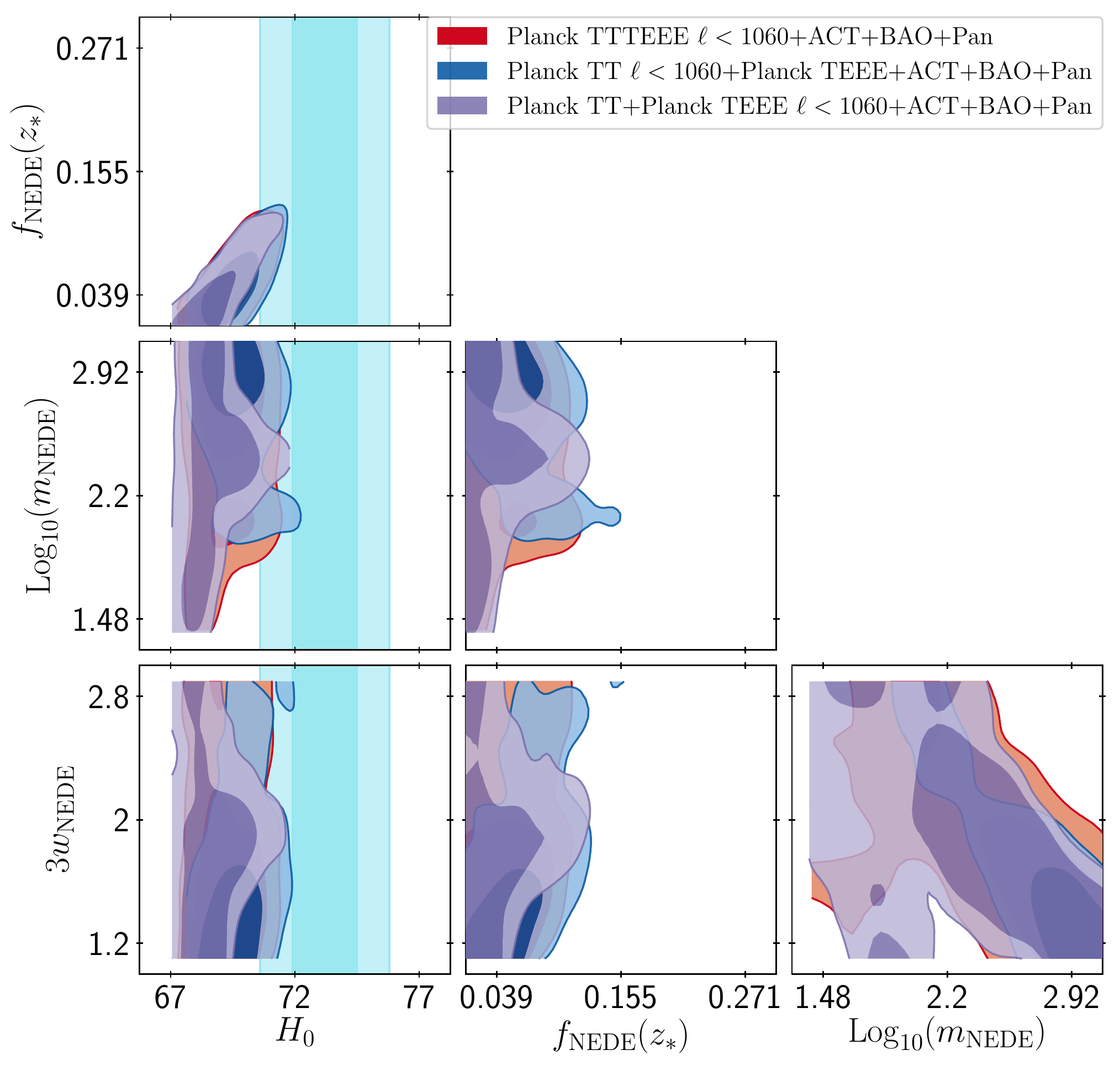}
    \caption{Same as Fig.~\ref{fig:axEDE_full_PlanckRes} for the NEDE case.}
    \label{fig:NEDE_full_PlanckRes}
\end{figure*}

The left panel of Figs.~\ref{fig:axEDE_full_PlanckRes} and \ref{fig:NEDE_full_PlanckRes} show the reconstructed axEDE and NEDE parameters respectively, when analyzing ACT together with either WMAP, {\it Planck} data restricted to $\ell<1060$, or the full multipole range of {\it Planck}. In the right panels, we show the difference between restricting {\it Planck} in TT, TEEE or both. All analyses also include BAO and Pantheon SN data. In the axEDE case, one can clearly see that high-$\ell$ TT data from {\it Planck} are driving the constraints. For NEDE however, the high-$\ell$ TEEE from {\it Planck} also kill the high-$H_0$/low trigger-mass mode. However, the high-$\ell$ TEEE data do allow for the high-mass mode with a relatively high $H_0\sim 70$ km/s/Mpc.

\section{NEDE model with $w=2/3$}
\label{app:NEDE_w23}
\begin{figure}
    \centering
    \includegraphics[width=\columnwidth]{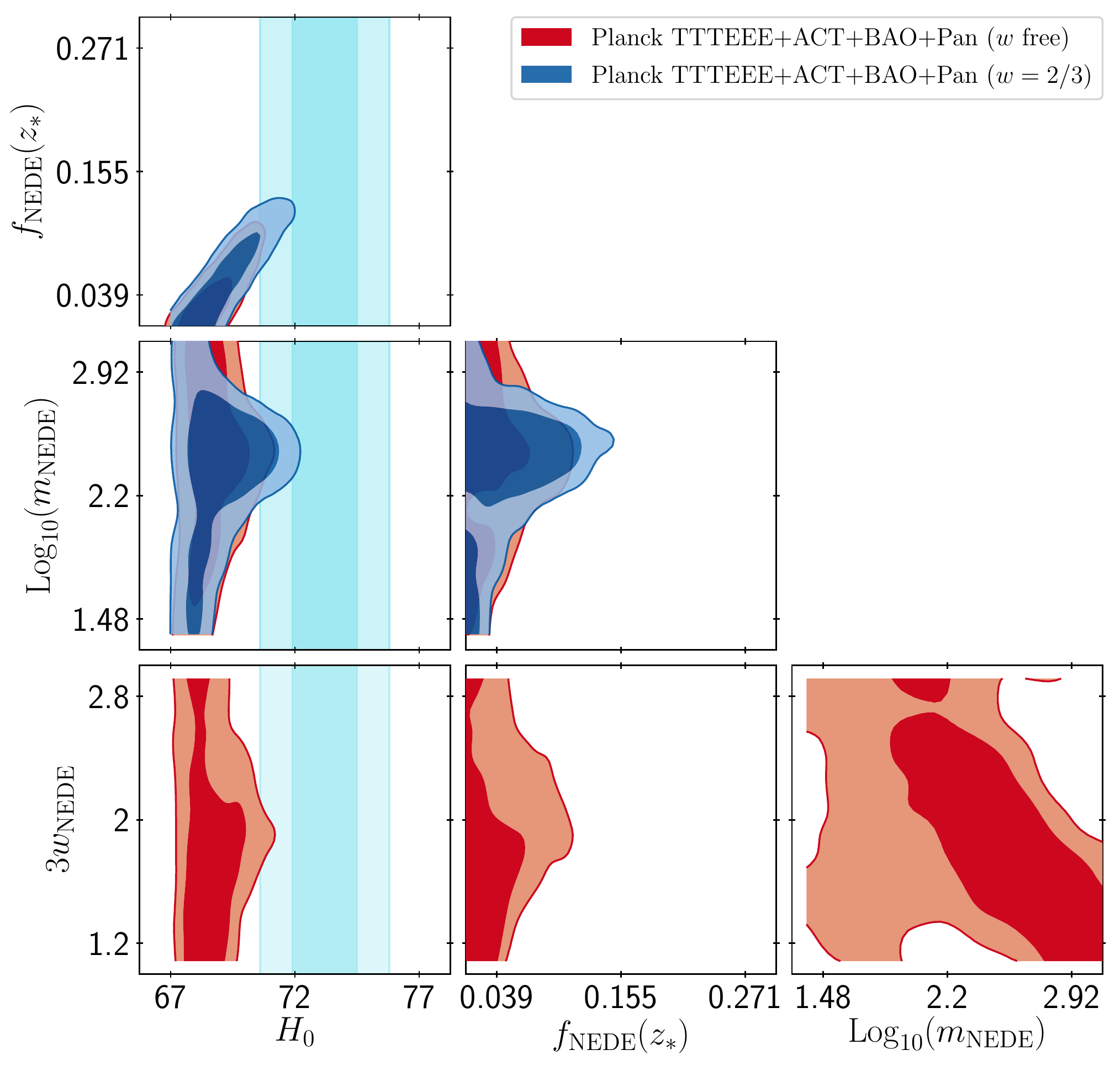}
    \caption{A comparison between the 2D posteriors reconstructed from a run against {\em Planck}+ACT+BAO+Pantheon when $w_{\rm NEDE}$ is let free to vary or fixed to $2/3$.}
    \label{fig:app_NEDE}
\end{figure}

In this study, we have kept the NEDE field equation of state $w_{\rm NEDE}$ free to vary. While we have shown that there is no clear preference emerging for specific values of $w_{\rm NEDE}$ from the analysis of WMAP and ACT data (see Tab.~\ref{tab:NEDE_WMAP}), it was noted in the literature that {\em Planck}+SH0ES favor $w_{\rm NEDE}=2/3$, such that previous studies have kept this parameter fixed in their baseline analyses \cite{Niedermann:2019olb,Niedermann:2020dwg,Niedermann:2020qbw}. Similarly here, when combining {\em Planck+}ACT+SH0ES, we find that the data favor $3w_{\rm NEDE}\simeq2\pm0.2$ (Tab.~\ref{tab:NEDE_Planck}). Yet, the tension between {\em Planck}+ACT and SH0ES only decreases to $2.9\sigma$. It is interesting to check whether the fact that the tension does not decrease much was due to complicated prior volume effects, and if keeping $w_{\rm NEDE}=2/3$ can alter conclusions regarding the NEDE model. We thus perform an additional run of the NEDE model against {\em Planck}+ACT+BAO+Pantheon with $w_{\rm NEDE}=2/3$.  We compare results with the $w_{\rm NEDE}$ free case in Fig.~\ref{fig:app_NEDE}. One can clearly see that $H_0$ can extend to higher values, and we find $H_0=69.02_{-1.5}^{+0.78}$ km/s/Mpc, a $\lesssim 1\sigma$ upward shift compared to the $w_{\rm NEDE}$ case. 
Moreover, in the $w_{\rm NEDE}=2/3$ case, we find $\chi^2_{\rm min}=4048.05$, while the $w_{\rm NEDE}$ free case led to $\chi^2_{\rm min}=4044.3$. Reducing the parameter space therefore only leads to a small change in $\chi^2_{\rm min}$.
To quantify tension, we compute the $Q_{\rm DMAP}$ metric between this run and the run that includes a prior on $M_b$ presented in Sec.~\ref{sec:Planck-vs-ACT}, for which letting $w_{\rm NEDE}$ free to vary led to a best-fit value of $2/3$.  We find  $Q_{\rm DMAP}(w=2/3)=2.1\sigma$, which is lower than in the $w_{\rm NEDE}$ free, but still much larger than the tension level in the axEDE model ($0.3\sigma$). We therefore conclude that our main conclusions are not changed by our choice of letting $w_{\rm NEDE}$ free to vary instead of keeping it fix to $2/3$.

\section{ACT vs ACT+WMAP+BAO+Pantheon best-fit in the axEDE cosmology}
\label{app:ACTvsWMAP}

We compare the residuals between EDE and $\Lambda$ for the best-fit cosmologies reconstructed with and without large-scale CMB information from WMAP, BAO and Pantheon data. One can see that the main impact of these data (and in particular WMAP), is to reduce the amplitude of the bump in EE and TT at $\ell\sim 400$. In a combined fit, we find that the fit to ACT data degrades slightly ($\sim+7$) in both cosmologies. Yet, the improvement provided by axEDE over $\Lambda$CDM within ACT stays the same ($\Delta\chi^2 \sim -10$), while the fit to WMAP is (marginally) improved as well ($\Delta\chi^2 \sim-3.5$); the other $\chi^2$s are relatively unaffected. As a result, the preference for non-zero $f_{\rm axEDE}(z_c)$ is slightly larger in the combined fit than in ACT only.

\section{$\chi^2_{\rm min}$ per experiment}
\label{app:chi2}

\begin{table*}
\scalebox{1}{
\begin{tabular}{|l|c|c|c|c|}
\hline
& $\Lambda$CDM & axEDE & NEDE, M1 & NEDE, M2 \\
\hline
ACTPol & 280.19 & 270.774 &271.82 &270.21  \\
$\tau$ prior & 0.01 & 0.13  &0.30 &0.00\\
\hline
total $\chi^2$  & 280.20 & 270.90  &272.12 &270.21 \\
\hline
\end{tabular}}
\caption{Best-fit $\chi^2$ per experiment (and total) for all models when fit to ACT alone.}
\label{tab:chi2_act}

\end{table*}

\begin{table*}
\scalebox{1}{
\begin{tabular}{|l|c|c|c|c|c|c|c|c|}
\hline
& \multicolumn{2}{c|}{$\Lambda$CDM} & \multicolumn{2}{c|}{axEDE} & \multicolumn{2}{c|}{NEDE, M1} & \multicolumn{2}{c|}{NEDE, M2} \\
\hline
ACTPol & 287.06 & 287.93 &276.04 &276.174 &276.39 &277.48 &272.43 &273.72 \\
WMAP & 5627.16& 5626.71 &5623.74 &5624.20 &5629.62 &5628.91 &5624.459 &5624.87 \\
Pantheon SNIa  & 1026.86 & 1026.92 &1026.71 &1026.68 &1026.69&1026.67 &1026.77 & 1026.77 \\
BAO~BOSS low$-z$ & 1.38 &1.81 &1.71 &1.94 &1.89 &2.20 &1.68 &1.84 \\
BAO BOSS DR12  & 3.85 &3.35 &3.45 &3.36 &3.31 &3.44 &3.30 &3.25 \\
$\tau$ prior & 0.001 & 0.02 &0.05 &0.001 &0.01&0.02 &0.001 &0.02\\
SH0ES  & -- & 17.29&-- &0.14 &-- &  0.18 &-- &4.09\\
\hline
total $\chi^2$  & 6946.31 & 6964.04&6931.70 &6932.50 &6937.92 &6938.90 &6928.65 &6934.55 \\
$\Delta \chi^2$  & 0 & 0 & -14.61 & -31.54 &-8.39 &-25.14 &-17.66 &-29.49 \\
\hline
$Q_{\rm DMAP}$ & \multicolumn{2}{c|}{$4.5\sigma$} &\multicolumn{2}{c|}{$0.3\sigma$} &  \multicolumn{2}{c|}{$1.0\sigma$} & \multicolumn{2}{c|}{$2.4\sigma$} \\
\hline
\end{tabular}}
\caption{Best-fit $\chi^2$ per experiment (and total) for all models when fit to WMAP and ACT.}
\label{tab:chi2_wmap}

\end{table*}

\begin{table*}
\scalebox{1}{
\begin{tabular}{|l|c|c|c|c|c|c|}
\hline
& \multicolumn{2}{c|}{$\Lambda$CDM} & \multicolumn{2}{c|}{axEDE} & \multicolumn{2}{c|}{NEDE} \\
\hline
ACTPol & 240.12 & 235.56 &235.56 &237.67 &235.20 & 239.05 \\
{\emph{Planck}}~high$-\ell$ TT,TE,EE & 2350.68 & 2350.63 & 2351.93 & 2347.98&2347.92 &2349.61   \\
{\emph{Planck}}~low$-\ell$ EE &396.28 &396.33 &395.81  &396.50 &397.55 &396.68 \\
{\emph{Planck}}~low$-\ell$ TT& 22.17 &  22.17&21.25  &20.74 &21.32 &20.918 \\
{\emph{Planck}}~lensing & 8.85 & 8.89 &9.95  &10.11 &9.76 &10.042 \\
Pantheon SN1a &1026.80 &  1026.07&1026.71 &1026.68 &1026.82 & 1026.73\\
BAO~BOSS low$-z$ & 1.44  & 1.41& 1.71  &2.19 &1.43 &1.88  \\
BAO BOSS DR12  & 3.79 & 3.86 &3.50  &3.48 &3.79 &3.40  \\
SH0ES  & -- &19.73 & --  &1.15 &-- &4.05\\
\hline
total $\chi^2$  & 4050.14 &  4070.47&4046.41  &4046.50 &4043.78 &4052.37 \\
\hline
\end{tabular}}
\caption{Best-fit $\chi^2$ per experiment (and total) for all models when fit to Planck and ACT.}
\label{tab:chi2_planck}

\end{table*}
\end{appendix}

\end{document}